\documentstyle[12pt,epsfig]{article}

\newcommand{\agt}{\,\rlap{\lower 3.5 pt \hbox{$\mathchar \sim$}} \raise 1pt
 \hbox {$>$}\,}
\newcommand{\alt}{\,\rlap{\lower 3.5 pt \hbox{$\mathchar \sim$}} \raise 1pt
 \hbox {$<$}\,}

\catcode`@=11
\newcount\@tempcntc
\def\@citex[#1]#2{\if@filesw\immediate\write\@auxout{\string\citation{#2}}\fi
  \@tempcnta\z@\@tempcntb\m@ne\def\@citea{}\@cite{\@for\@citeb:=#2\do
    {\@ifundefined
       {b@\@citeb}{\@citeo\@tempcntb\m@ne\@citea\def\@citea{,}{\bf ?}\@warning
       {Citation `\@citeb' on page \thepage \space undefined}}%
    {\setbox\z@\hbox{\global\@tempcntc0\csname b@\@citeb\endcsname\relax}%
     \ifnum\@tempcntc=\z@ \@citeo\@tempcntb\m@ne
       \@citea\def\@citea{,}\hbox{\csname b@\@citeb\endcsname}%
     \else
      \advance\@tempcntb\@ne
      \ifnum\@tempcntb=\@tempcntc
      \else\advance\@tempcntb\m@ne\@citeo
      \@tempcnta\@tempcntc\@tempcntb\@tempcntc\fi\fi}}\@citeo}{#1}}
\def\@citeo{\ifnum\@tempcnta>\@tempcntb\else\@citea\def\@citea{,}%
  \ifnum\@tempcnta=\@tempcntb\the\@tempcnta\else
   {\advance\@tempcnta\@ne\ifnum\@tempcnta=\@tempcntb \else \def\@citea{--}\fi
    \advance\@tempcnta\m@ne\the\@tempcnta\@citea\the\@tempcntb}\fi\fi}
\catcode`@=12
\begin{document}
\title{\vskip-3cm{\baselineskip14pt
\centerline{\normalsize DESY 04-023\hfill ISSN 0418-9833}
\centerline{\normalsize NSF-KITP-04-24\hfill}
\centerline{\normalsize hep-ph/0404039\hfill}
\centerline{\normalsize April 2004\hfill}}
\vskip1.5cm
Associated Production of Bottomonia and Higgs Bosons at Hadron Colliders}
\author{{\sc Bernd A. Kniehl}\thanks{Permanent address: II. Institut f\"ur
Theoretische Physik, Universit\"at Hamburg, Luruper Chaussee 149, 22761
Hamburg, Germany.}\\
{\normalsize Kavli Institute for Theoretical Physics, University of
California,}\\
{\normalsize Santa Barbara, CA 93106--4030, USA}\\
\\
{\sc Caesar P. Palisoc}\\
{\normalsize National Institute of Physics, University of the Philippines,}\\
{\normalsize Diliman, Quezon City 1101, Philippines}\\
\\
{\sc Lennart Zwirner}\\
{\normalsize II. Institut f\"ur Theoretische Physik, Universit\"at Hamburg,}\\
{\normalsize Luruper Chaussee 149, 22761 Hamburg, Germany}}

\date{}

\maketitle

\thispagestyle{empty}

\begin{abstract}
We study the associated production of bottomonia and Higgs bosons at hadron
colliders within the factorization formalism of nonrelativistic quantum
chromodynamics providing all contributing partonic cross sections in analytic
form.
While such processes tend to be suppressed in the standard model, they may
have interesting cross sections in its minimal supersymmetric extension,
especially at large values of $\tan\beta$, where the bottom Yukawa couplings
are enhanced.
We present numerical results for the processes involving the lighter $CP$-even
$h^0$ boson and the $CP$-odd $A^0$ boson appropriate for the Fermilab Tevatron
and the CERN LHC.

\medskip

\end{abstract}

\newpage

\section{Introduction}
\label{sec:one}

Since the discovery of the $J/\psi$ meson in 1974, heavy quarkonium has
provided a useful laboratory for quantitative tests of quantum chromodynamics
(QCD) and, in particular, of the interplay of perturbative and nonperturbative
phenomena.
The factorization formalism of nonrelativistic QCD (NRQCD) \cite{bbl} provides
a rigorous theoretical framework for the description of heavy-quarkonium
production and decay.
This formalism implies a separation of short-distance coefficients, which can 
be calculated perturbatively as expansions in the strong-coupling constant
$\alpha_s$, from long-distance matrix elements (MEs), which must be extracted
from experiment.
The relative importance of the latter can be estimated by means of velocity
scaling rules; i.e., the MEs are predicted to scale with a definite power of
the heavy-quark ($Q$) velocity $v$ in the limit $v\ll1$.
In this way, the theoretical predictions are organized as double expansions in
$\alpha_s$ and $v$.
A crucial feature of this formalism is that it takes into account the complete
structure of the $Q\overline{Q}$ Fock space, which is spanned by the states
$n={}^{2S+1}L_J^{(a)}$ with definite spin $S$, orbital angular momentum
$L$, total angular momentum $J$, and color multiplicity $a=1,8$.
In particular, this formalism predicts the existence of color-octet (CO)
processes in nature.
This means that $Q\overline{Q}$ pairs are produced at short distances in CO
states and subsequently evolve into physical, color-singlet (CS) quarkonia by
the nonperturbative emission of soft gluons.
In the limit $v\to0$, the traditional CS model (CSM) \cite{ber} is recovered.
The greatest triumph of this formalism was that it was able to correctly 
describe \cite{bra,cho1,cho2} the cross section of inclusive charmonium
hadroproduction measured in $p\overline{p}$ collisions at the Fermilab
Tevatron \cite{abe}, which had turned out to be more than one order of
magnitude in excess of the theoretical prediction based on the CSM.

Apart from this phenomenological drawback, the CSM also suffers from severe
conceptual problems indicating that it is incomplete.
These include the presence of logarithmic infrared divergences in the
${\cal O}(\alpha_s)$ corrections to $P$-wave decays to light hadrons and in
the relativistic corrections to $S$-wave annihilation \cite{bar}, and the lack
of a general argument for its validity in higher orders of perturbation
theory.
While the $k_T$-factorization \cite{sri} and hard-comover-scattering
\cite{hoy} approaches manage to bring the CSM prediction much closer to the
Tevatron data, they do not cure the conceptual defects of the CSM.
The color evaporation model \cite{cem}, which is intuitive and useful for
qualitative studies, also significantly improves the description of the
Tevatron data as compared to the CSM \cite{sch}.
However, it does not account for the process-specific weights of the CS and CO
contributions, but rather assumes a fixed ratio of $1:7$.
In this sense, a coequal alternative to the NRQCD factorization formalism is 
presently not available.

In order to convincingly establish the phenomenological significance of the CO
processes, it is indispensable to identify them in other kinds of high-energy
experiments as well.
Studies of charmonium production in $ep$ photoproduction, $ep$ and $\nu N$
deep-inelastic scattering (DIS), $e^+e^-$ annihilation, $\gamma\gamma$
collisions, and $b$-hadron decays may be found in the literature; see
Refs.~\cite{fle,kra} and references cited therein.
Furthermore, the polarization of $\psi^\prime$ mesons produced directly
\cite{ben} and of $J/\psi$ mesons produced promptly \cite{bkl,kni}, i.e.,
either directly or via the feed-down from heavier charmonia, which also
provides a sensitive probe of CO processes, was investigated.
Until recently, none of these studies was able to prove or disprove the NRQCD
factorization hypothesis.
However, H1 data of $ep\to eJ/\psi+X$ in DIS at HERA \cite{h1} and DELPHI data
of $\gamma\gamma\to J/\psi+X$ at LEP2 \cite{delphi} now provide first
independent evidence for it \cite{ep,gg}.

Recently, we studied the associated production of charmonia and electroweak 
bosons at present and future colliders, including the Tevatron (Run~II), the
CERN Large Hadron Collider (LHC), the DESY TeV-Energy Superconducting Linear
Accelerator (TESLA), and the DESY electron-proton supercollider THERA, which
uses the lepton beam of TESLA and the proton beam of the DESY Hadron-Electron
Ring Accelerator (HERA) \cite{jb}.
Our assessment of the observability of the various processes in the various 
experiments can be summarized as follows: 
the processes with a photon will abundantly take place in all considered
experiments;
the processes with a $Z$ or $W$ boson will produce considerable yields at the
hadron colliders, namely, several hundred (ten thousand) events per year at the
Tevatron (LHC), while they significantly fall short of the one-event-per-year
mark at TESLA and THERA;
the processes with a standard model (SM) Higgs boson ($\cal H$) are predicted
to be too rare to be observable in any of the considered experiments.
As for $J/\psi+W$ associated production at the Tevatron, the conclusions that
had been reached in Ref.~\cite{phi} concur with ours \cite{jb}.
The associated production of bottomonia and $Z$ or $W$ bosons at the Tevatron
and the LHC was investigated in Ref.~\cite{lee}, with the conclusion that such
events represent a challenge at the Tevatron, while they should certainly be
observable at the LHC.
The CDF Collaboration \cite{aco} did not find evidence for $\Upsilon(1S)+Z$ or
$\Upsilon(1S)+W^\pm$ signals in Run~I at the Tevatron, but established upper
bounds on the respective production cross sections, which are well compatible
with the theoretical predictions of Ref.~\cite{lee}.
However, with the expected Run~II increase of integrated luminosity and the 
upgrades of the CDF detector, they expect to achieve a sensitivity sufficient
to observe these signals \cite{aco}.
The cross sections of all partonic processes contributing to the associated 
production of heavy quarkonia, with
${}^{2S+1}L_J={}^1\!S_0,{}^3\!S_1,{}^1\!P_1,{}^3\!P_J$ ($J=0,1,2$), and
photons, $Z$ bosons, and $W$ bosons in photon-photon, photon-hadron, and
hadron-hadron collisions may be found in Ref.~\cite{jb}.

In this paper, we take the next step and consider the production of
bottomonia in association with Higgs bosons concentrating on hadron-hadron 
collisions.
Guided by our previous findings in connection with charmonium \cite{jb}, we 
expect such signals to be below the edge of observability in two-photon
collisions at TESLA and photon-hadron collisions at THERA.
Therefore, we do not include the latter production mechanisms in our present
phenomenological analysis.
However, the relevant partonic cross sections may be obtained from our
analytic results by adjusting overall color factors and coupling constants,
as is explained in the Appendix.

In the SM, the total cross section of $pp\to J/\psi+{\cal H}$, with
Higgs-boson mass $m_{\cal H}=115$~GeV, at the LHC was predicted to be
$2.0\times10^{-2}$~fb, which corresponds to 4 signal events per year \cite{jb}.
Taking into account the branching fractions of the $J/\psi$ and $\cal H$
decays to the detected particles and the acceptance cuts required for
background suppression, it becomes apparent that the production rate would
need to be increased by a few orders of magnitude before a sound signal could
be observed.
Since the SM Yukawa couplings are proportional to the quark masses, passing 
from charmonium to bottomonium yields an enhancement by a factor of
$m_b^2/m_c^2\approx10$.
Moreover, the MEs of bottomonium \cite{kra,eic} are typically one order of 
magnitude larger than their counterparts for charmonium \cite{bkl}, with the
exception of $\langle{\cal O}^{\Upsilon(nS)}[{}^1\!S_0^{(8)}]\rangle$ and
$\langle{\cal O}^{\Upsilon(nS)}[{}^3\!P_J^{(8)}]\rangle$, which tend to be
comparable to their counterparts for $\psi(nS)$.
On the other hand, the dynamical quark-mass dependence acts in the opposite
direction.
In fact, in the case of the total cross section of $pp\to J/\psi+{\cal H}$,
with $m_{\cal H}=115$~GeV at the LHC, replacing $m_c$ by $m_b$ while keeping
the Yukawa coupling and the MEs fixed leads to a decrease by a bit more than 
one order of magnitude, so that the Yukawa coupling enhancement is slightly
overtaken.
Detailed calculation reveals that the total cross section of
$pp\to \Upsilon(1S)+{\cal H}$ at the LHC exceeds the one of
$pp\to J/\psi+{\cal H}$ by a factor of 3.6, the result being
$7.3\times10^{-2}$~fb, which corresponds to 15 signal events per year. 

The situation should be considerably more promising in the context of
supersymmetric theories, where the Yukawa couplings depend on additional input
parameters and may be significantly enhanced relative to the reference values
in the SM if nature has chosen these parameters accordingly.
In the following, we concentrate on the most popular of such theories, the
minimal supersymmetric extension of the SM (MSSM).
The Higgs sector of the MSSM consists of a two-Higgs-doublet model of type~II
and accommodates a quintet of physical Higgs bosons: the neutral $CP$-even $h$
and $H$ bosons, the neutral $CP$-odd $A$ boson, and the charged $H^\pm$-boson
pair.
At the tree level, the MSSM Higgs sector has two free parameters, which are
usually taken to be the mass $m_A$ of the $A$ boson and the ratio
$\tan\beta=v_2/v_1$ of the vacuum expectation values of the two Higgs
doublets.
The masses $m_h$, $m_H$, and $m_{H^\pm}$ of the $h$, $H$, and $H^\pm$ bosons,
respectively, and the mixing angle $\alpha$ that rotates the weak $CP$-even
Higgs eigenstates into the mass eigenstates $h$ and $H$ are then functions of
$m_A$ and $\tan\beta$.
As is well known, these functions receive significant electroweak radiative
corrections, whose leading terms are of ${\cal O}\left(G_Fm_t^4/m_Z^2\right)$,
where $G_F$ is Fermi's constant, $m_t$ is the top-quark mass, and $m_Z$ is the
$Z$-boson mass, and must not be neglected; for a review, see Ref.~\cite{spi}.
For the sake of our exploratory study, it is sufficient to make a few 
simplifying assumptions in the treatment of these corrections.
Specifically, we neglect subleading effects due to nonzero values of the
Higgs-higgsino mass parameter $\mu$ and the trilinear Higgs-sfermion couplings
$A_t$ and $A_b$.
The leading radiative corrections can then be parameterized by the quantity
\begin{equation}
\epsilon=\frac{3G_Fm_t^4}{\pi^2\sqrt2\sin^2\beta}
\ln\left(1+\frac{m_S^2}{m_t^2}\right),
\end{equation}
where $m_S$ is the common squark mass, and we have
\begin{eqnarray}
m_h^2&=&\frac{1}{2}\left[m_A^2+m_Z^2+\epsilon
\vphantom{\sqrt{\left(m_A^2+m_Z^2+\epsilon\right)^2-4m_A^2m_Z^2\cos^2(2\beta)
-4\epsilon\left(m_A^2\sin^2\beta+m_Z^2\cos^2\beta\right)}}
\right.
\nonumber\\
&&{}-\left.
\sqrt{\left(m_A^2+m_Z^2+\epsilon\right)^2-4m_A^2m_Z^2\cos^2(2\beta)
-4\epsilon\left(m_A^2\sin^2\beta+m_Z^2\cos^2\beta\right)}\right],
\label{eq:mh}\\
m_H^2&=&m_A^2+m_Z^2-m_h^2+\epsilon,
\label{eq:mH}\\
m_{H^\pm}^2&=&m_A^2+m_W^2,\\
\tan(2\alpha)&=&\frac{\left(m_A^2+m_Z^2\right)\sin(2\beta)}
{\left(m_A^2-m_Z^2\right)\cos(2\beta)+\epsilon},
\label{eq:al}
\end{eqnarray}
where $m_W$ is the $W$-boson mass.
Solving Eq.~(\ref{eq:al}) for $\alpha$, one needs to select the branch with
$-\pi/2<\alpha<0$.
When $m_h$ and $\tan\beta$ are chosen to be the input parameters, then
Eq.~(\ref{eq:mh}) can be solved for $m_A$, which leads to
\begin{equation}
m_A^2=\frac{m_h^2(m_Z^2-m_h^2)+\epsilon\left(m_h^2-m_Z^2\cos^2\beta\right)}
{m_Z^2\cos^2(2\beta)-m_h^2+\epsilon\sin^2\beta},
\label{eq:mA}
\end{equation}
and $\alpha$ can be obtained from Eq.~(\ref{eq:al}) as before.
From Eq.~(\ref{eq:mA}), we glean that the upper bound on $m_h^2$ is shifted
from its tree-level value, $m_Z^2\cos^2(2\beta)$, to 
\begin{equation}
m_h^2<m_Z^2\cos^2(2\beta)+\epsilon\sin^2\beta.
\label{eq:max}
\end{equation}
Combining Eqs.~(\ref{eq:mh}) and (\ref{eq:mH}) and noticing that the radicant
in Eq.~(\ref{eq:mh}) can be written as
$\left[m_A^2-m_Z^2\cos(4\beta)+\epsilon\cos(2\beta)\right]^2
+\left[m_Z^2\sin(4\beta)-\epsilon\sin(2\beta)\right]^2$,
we learn that
\begin{equation}
m_H^2>m_h^2+\left|m_Z^2\sin(4\beta)-\epsilon\sin(2\beta)\right|.
\end{equation}
Furthermore, inserting Eq.~(\ref{eq:max}) into Eq.~(\ref{eq:mH}), we find that
\begin{equation}
m_H^2>m_A^2+m_Z^2\sin^2(2\beta)+\epsilon\cos^2\beta.
\end{equation}
In other words, $H$ is always heavier than $h$ and $A$.

If we assume only the first four quark flavors to be active in the colliding
hadrons, then the associated production of bottomonium and a charged Higgs
boson is suppressed by the smallness of the $\overline{d}uH^-$ and
$\overline{s}cH^-$ couplings.
If we also take bottom to be an active flavor, then the final state must 
contain an additional top quark, $W$ boson, or charged Higgs boson, which
substantially squeeze the available phase space and lead to distinctive
signals.
For these reasons, we only consider the neutral MSSM Higgs bosons $\Phi=h,H,A$
in the following.
The $q\overline{q}\Phi$ couplings $y_q^\Phi$ emerge by scaling the SM Yukawa
couplings with factors $g_q^\Phi$; i.e.,
\begin{equation}
y_q^\Phi=2^{1/4}G_F^{1/2}m_qg_q^\Phi.
\label{eq:yq}
\end{equation}
The values of $g_q^\Phi$ are specified in Table~\ref{tab:gy}.
From Table~\ref{tab:gy}, we observe that the values of $g_b^\Phi$ rise
linearly with $\tan\beta$ if $\tan\beta\gg1$.
In order to obtain sufficiently large cross sections, we are particularly
interested in light Higgs bosons with strong couplings to the $b$ quark.
We thus focus our attention on the $h$ and $A$ bosons, and on the
large-$\tan\beta$ regime.

\begin{table}[ht]
\begin{center}
\caption{Values of $g_q^\Phi$ in Eq.~(\ref{eq:yq}) for generic up and down
quarks, $U$ and $D$, respectively.}
\label{tab:gy}
\medskip
\begin{tabular}{|c|ccc|}
\hline\hline
$q$ & $g_q^h$ & $g_q^H$ & $g_q^A$ \\
\hline
$U$ & $\cos\alpha/\sin\beta$ & $\sin\alpha/\sin\beta$ & $\cot\beta$ \\
$D$ & $-\sin\alpha/\cos\beta$ & $\cos\alpha/\cos\beta$ & $\tan\beta$ \\
\hline\hline
\end{tabular}
\end{center}
\end{table}

This paper is organized as follows.
In Sec.~\ref{sec:two}, we present our analytic results and explain how to
evaluate the cross sections of the associated production of bottomonia and
Higgs bosons in hadronic collisions.
The contributing partonic cross sections are collected in the Appendix.
In Sec.~\ref{sec:three}, we present our numerical results.
Our conclusions are summarized in Sec.~\ref{sec:four}.

\section{Analytic results}
\label{sec:two}

In this section, we present our analytic results for the cross sections of the
reactions $AB\to CD$, where $A$ and $B$ are the incoming hadrons, $C$ is a
heavy-quarkonium state, with
${}^{2S+1}L_J={}^1\!S_0,{}^3\!S_1,{}^1\!P_1,{}^3\!P_J$ ($J=0,1,2$), and
$D=h,A$.
The results for $D={\cal H},H$ may be obtained from those for $D=h$ by
accordingly replacing $y_Q^h$ and $m_h$.
We also list the formulas for the cases where $A$ and/or $B$ are real or
quasireal photons, appropriate for two-photon and photon-hadron collisions,
respectively.

We work in the fixed-flavor-number scheme; i.e., $A$ and $B$ contain $n_f=4$
active quark flavors $q=u,d,s,c$ if $Q=b$ and $n_f=3$ active quark flavors
$q=u,d,s$ if $Q=c$.
As required by parton-model kinematics, we treat the quark flavors $q$ as
massless.
The quark flavor $Q$, with mass $m_Q$, only appears in the final state.
The $b\overline{b}$ Fock states contributing at LO in $v$ are
$n={}^1\!S_0^{(1)},{}^1\!S_0^{(8)},{}^3\!S_1^{(8)},{}^1\!P_1^{(8)}$ if
$H=\eta_b(nS)$;
$n={}^3\!S_1^{(1)},{}^1\!S_0^{(8)},{}^3\!S_1^{(8)},{}^3\!P_J^{(8)}$ if
$H=\Upsilon(nS)$;
$n={}^1\!P_1^{(1)},{}^1\!S_0^{(8)}$ if $H=h_b(nP)$; and
$n={}^3\!P_J^{(1)},{}^3\!S_1^{(8)}$ if $H=\chi_{bJ}(nP)$, where $J=0,1,2$.
Their MEs satisfy the multiplicity relations
\begin{eqnarray}
\left\langle{\cal O}^{\Upsilon(nS)}\left[{}^3\!P_J^{(8)}\right]\right\rangle
&=&(2J+1)
\left\langle{\cal O}^{\Upsilon(nS)}\left[{}^3\!P_0^{(8)}\right]\right\rangle,
\nonumber\\
\left\langle{\cal O}^{\chi_{bJ}(nP)}\left[{}^3\!P_J^{(1)}\right]\right\rangle
&=&(2J+1)
\left\langle{\cal O}^{\chi_{b0}(nP)}\left[{}^3\!P_0^{(1)}\right]\right\rangle,
\nonumber\\
\left\langle{\cal O}^{\chi_{bJ}(nP)}\left[{}^3\!S_1^{(8)}\right]\right\rangle
&=&(2J+1)
\left\langle{\cal O}^{\chi_{b0}(nP)}\left[{}^3\!S_1^{(8)}\right]\right\rangle,
\label{eq:mul}
\end{eqnarray}
which follow to LO in $v$ from heavy-quark spin symmetry.
The CS MEs of $\Upsilon(nS)$ and $\chi_{b0}(nP)$ are related to the respective
radial wave functions at the origin as
\begin{eqnarray}
\left\langle{\cal O}^{\Upsilon(nS)}\left[{}^3\!S_1^{(1)}\right]\right\rangle
=\frac{9}{2\pi}|R_{nS}(0)|^2,
\nonumber\\
\left\langle{\cal O}^{\chi_{b0}(nP)}\left[{}^3\!P_0^{(1)}\right]\right\rangle
=\frac{9}{2\pi}\left|R_{nP}^\prime(0)\right|^2,
\label{eq:wf}
\end{eqnarray}
 respectively.
In our numerical analysis, we only include the $\Upsilon(1S)$ and
$\chi_{bJ}(1P)$ mesons.
For completeness and future use, we also list formulas for all the other
bottomonia.
The assignments for the various charmonia are analogous.

We now turn to the partonic subprocesses $ab\to Q\overline{Q}[n]D$.
The differential cross section of such a process is calculated from the
pertaining transition-matrix element ${\cal T}$ as
$d\sigma/dt=\overline{|{\cal T}|^2}/(16\pi s^2)$, where the average is over
the spin and color degrees of freedom of $a$ and $b$ and the spin of $D$ is
summed over.
We apply the covariant-projector method of Ref.~\cite{pet} to implement the
$Q\overline{Q}$ Fock states $n$ according to the NRQCD factorization
formalism.

The following partonic subprocesses contribute to LO in $\alpha_s$ and $v$:
\begin{eqnarray}
gg&\to&Q\overline{Q}[\varsigma^{(1)}]D,
\label{eq:gg}\\
gg&\to&Q\overline{Q}[\varsigma^{(8)}]D,
\label{eq:co}\\
q\overline{q}&\to&Q\overline{Q}[\varsigma^{(8)}]D,
\label{eq:qq}\\
\gamma g&\to&Q\overline{Q}[\varsigma^{(8)}]D,
\label{eq:pg}\\
\gamma\gamma&\to&Q\overline{Q}[\varsigma^{(1)}]D,
\label{eq:pp}
\end{eqnarray}
where $\varsigma={}^1\!S_0,{}^3\!S_1,{}^1\!P_1,{}^3\!P_J$ with $J=0,1,2$.
For the reason explained above, $q=c$ must not be included if $Q=c$.
The processes $\gamma g\to Q\overline{Q}[\varsigma^{(1)}]D$ and
$\gamma\gamma\to Q\overline{Q}[\varsigma^{(8)}]D$ are forbidden by color
conservation.
Furthermore, the processes $q\overline{q}\to Q\overline{Q}[\varsigma^{(1)}]D$
are prohibited because the $Q$-quark line is connected with the $q$-quark line
by one gluon, which transmits color to the $Q\overline{Q}$ pair.
Finally, due to charge-conjugation invariance, the processes
$gg\to Q\overline{Q}[\varsigma^{(1)}]D$,
$\gamma g\to Q\overline{Q}[\varsigma^{(8)}]D$, and
$\gamma\gamma\to Q\overline{Q}[\varsigma^{(1)}]D$ are forbidden for
$\varsigma={}^3\!S_1,{}^1\!P_1$, and the processes
$q\overline{q}\to Q\overline{Q}[\varsigma^{(8)}]D$ are forbidden for
$\varsigma={}^1\!S_0,{}^3\!P_J$.

The differential cross sections $d\sigma/dt$ of processes
(\ref{eq:gg})--(\ref{eq:pp}) are listed in the Appendix.
We combine the results proportional to the CO MEs
$\left\langle{\cal O}^{\Upsilon(nS)}\left[{}^3\!P_J^{(8)}\right]\right\rangle$
and\break
$\left\langle{\cal O}^{\chi_{bJ}(nP)}\left[{}^3\!S_1^{(8)}\right]\right
\rangle$,
for fixed value of the principal quantum number $n$,
exploiting the multiplicity relations of Eq.~({\ref{eq:mul}).
The results for processes (\ref{eq:pg}) and (\ref{eq:pp}) are obtained from
those for processes (\ref{eq:gg}) by adjusting overall color factors and
coupling constants as explained in the Appendix.
Similar relations also exist between processes (\ref{eq:gg}) and (\ref{eq:co})
for $\varsigma={}^1\!S_0,{}^3\!P_J$.

The evaluation of the hadronic cross sections and their distributions proceeds
as explained in Ref.~\cite{jb}.
Specifically, we consider the distributions in the transverse momentum $p_T$
common to $C$ and $D$, the rapidities $y_C$ and $y_D$ of $C$ and $D$, 
respectively, and the $CD$ invariant mass $m_{CD}$.

\section{Numerical results}
\label{sec:three}

We are now in a position to explore the phenomenological consequences of our
calculations.
We focus our attention on the cases $C=\Upsilon(1S),\chi_{bJ}(1P)$.
These bottomonia can be efficiently identified experimentally, and their MEs
are relatively well constrained \cite{kra,eic}.
The predicted cross-section distributions for the $\Upsilon(2S)$ and
$\Upsilon(3S)$ mesons are similar to those for the $\Upsilon(1S)$ meson, but
their normalizations are somewhat suppressed due to smaller MEs
\cite{kra,eic}.
The MEs of the $\chi_{bJ}(2P)$ mesons are very similar to those of the
$\chi_{bJ}(1P)$ mesons \cite{kra,eic}, and so are their predicted
cross-section distributions.
The $\eta_b(1S)$ meson needs experimental confirmation, while no events for
$\eta_b(nS)$ mesons with $n>1$ and $h_b(nS)$ mesons have been seen so far
\cite{pdg}.

We first describe our theoretical input and the kinematic conditions.
We use $m_b=4.88$~GeV \cite{kra,eic}, $m_t=174.3$~GeV, $m_Z=91.1876$~GeV,
$G_F=1.16639\times10^{-5}$~GeV${}^{-2}$, and the LO formula for
$\alpha_s^{(n_f)}(\mu_r)$ \cite{pdg} with $n_f=4$ and asymptotic scale 
parameter $\Lambda_{\rm QCD}^{(4)}=192$~MeV \cite{cteq}.
As for the proton PDFs, we use the LO set from the CTEQ Collaboration (CTEQ5L)
\cite{cteq}.
Unless otherwise specified, we choose the renormalization and factorization
scales to be $\mu_r=\mu_f=\sqrt{m_T^Cm_T^D}$, where $m_T^C=\sqrt{m_C^2+p_T^2}$
is the transverse mass of $C$ and similarly for $D$.
We evaluate
$\left\langle{\cal O}^{\Upsilon(1S)}\left[{}^3\!S_1^{(1)}\right]\right\rangle$
and
$\left\langle{\cal O}^{\chi_{b0}(1P)}\left[{}^3\!P_0^{(1)}\right]\right\rangle$
from Eq.~(\ref{eq:wf}) with the values of $|R_{1S}(0)|^2$ and
$\left|R_{1P}^\prime(0)\right|^2$ determined in Ref.~\cite{eic} using the
QCD-motivated potential by Buchm\"uller and Tye \cite{buc}.
As for the $\Upsilon(1S)$ and $\chi_{bJ}(1P)$ CO MEs, we adopt as our default
set the one determined in Ref.~\cite{kra} through a fit to recent CDF data
\cite{dac} using the same proton PDFs and CS MEs as we do.
Our MSSM benchmark scenario is defined by the parameter set $\tan\beta=50$,
$m_h=100$~GeV or $m_A=100$~GeV, and $m_S=1$~TeV, and is well inside the
currently allowed region of the MSSM parameter space \cite{pdg}.
We then vary $\tan\beta$, $m_h$, and $m_A$ one at a time in the ranges
$2<\tan\beta<60$, $90<m_h<128$~GeV, and $90<m_A<500$~GeV, respectively.
If $D=h$, then we take $m_h$ as an input parameter and evaluate $m_A$ from
Eq.~(\ref{eq:mA}).
The hadronic center-of-mass energy is $\sqrt S=2$~TeV in Run~II at the
Tevatron and $\sqrt S=14$~TeV at the LHC.

Since our study is at an exploratory level, we refrain from presenting a
full-fledged quantitative estimate of the theoretical uncertainties in our
predictions.
Experience from previous analyses of charmonium production within the NRQCD
factorization formalism \cite{ep,gg,nun} leads us to expect relative errors of
the order of $\pm50\%$.
This error estimate should be on the conservative side, since the convergence
property of the NRQCD perturbative expansion is expected to be considerably
improved as one passes from charmonium to bottomonium, thanks to the reduction
in size of $\alpha_s$ and $v$.
To be specific, we assess the theoretical uncertainties arising from the lack
of knowledge of the precise values of the bottomonium MEs and from the freedom
in the choice of the renormalization and factorization scales.

We are now in a position to present our numerical results.
Figures~\ref{fig:l} and \ref{fig:t} are devoted to $pp\to CD+X$ at the LHC and
to $p\overline{p}\to CD+X$ in Run~II at the Tevatron, respectively.
In each figure, part (a) gives the $p_T$ distributions $d\sigma/dp_T$ (upper
panel) and the $m_{CD}$ distributions $d\sigma/dm_{CD}$ (lower panel),
part (b) the $y_C$ distributions $d\sigma/dy_C$ (upper panel) and the $y_D$
distributions $d\sigma/dy_D$ (lower panel), and part (c) the total cross
sections $\sigma$ as functions of $\tan\beta$ (upper panel) and $m_D$ (lower
panel).
In each part, the left column refers to $D=h$ and the right one to $D=A$.
In each frame, we separately consider $C=\Upsilon(1S),\chi_{cJ}(1P)$, both in
NRQCD and the CSM.
It is summed over $C=\chi_{c0}(1P),\chi_{c1}(1P),\chi_{c2}(1P)$.
In the following, $n=1$ is implied whenever the label $(nL)$ is omitted.

As is evident from Sec.~\ref{sec:two}, $\Upsilon+h$ and $\Upsilon+A$
associated production proceeds through process (\ref{eq:co}) with
$\varsigma={}^1\!S_0,{}^3\!S_1,{}^3\!P_J$ and process (\ref{eq:qq})
with $\varsigma={}^3\!S_1$ (see solid lines in Figs.~\ref{fig:l} and 
\ref{fig:t}) and is forbidden to LO in the CSM.
On the other hand, $\chi_{bJ}+h$ and $\chi_{bJ}+A$ associated production is
already possible in the CSM, through process (\ref{eq:gg}) with
$\varsigma={}^3\!P_J$ (see dotted lines in Figs.~\ref{fig:l} and \ref{fig:t});
in NRQCD, also the CO processes (\ref{eq:co}) and (\ref{eq:qq}) with
$\varsigma={}^3\!S_1$ contribute.

In the study of the associated production of heavy quarkonia and electroweak
gauge bosons \cite{jb}, dominant contributions at $p_T\gg m_C$ were found to
generally arise from so-called {\it fragmentation-prone} partonic subprocesses,
which contain a gluon with small virtuality, $q^2=m_C^2$, that splits into a
$Q\overline{Q}$ pair in the Fock state $n={}^3\!S_1^{(8)}$.
Such processes are absent here because the Higgs bosons are always radiated 
from the outgoing $Q$ or $\overline{Q}$ quark lines before the latter form an
asymptotic Fock state $n$.

We now turn to the predictions for the LHC.
From Figs.~\ref{fig:l}(a)--(c), we observe that there is hardly any difference
between the various distributions for $D=h$ and their counterparts for $D=A$,
except for a trivial difference in the $m_h$ and $m_A$ dependences, which is
due to the upper bound on $m_h$, at $m_h\approx128$~GeV.
In each frame, the three curves exhibit similar shapes (except for the $p_T$
distributions), but significantly differ in their overall normalizations.
Within NRQCD, the results for $C=\chi_{bJ}$ typically exceed those for
$C=\Upsilon$ by a factor of 2.
As for $C=\chi_{bJ}$, the CSM predictions fall short of the NRQCD ones by a
factor of 3 to 4.
From Fig.~\ref{fig:l}(a), we read off that the $p_T$ and $m_{CD}$
distributions peak at $p_T\approx6$~GeV and $m_{CD}\approx116$~GeV, 
respectively.
Beyond their peaks, the CSM predictions for $C=\chi_{bJ}$ fall off more
rapidly with increasing value of $p_T$ than the NRQCD ones.
The $y_C$ and $y_D$ distributions in Fig.~\ref{fig:l}(b) are symmetric about
the origin, reflecting the symmetry of the experimental set-up, and their
maxima are rather broad.
Figure~\ref{fig:l}(c) displays the quadratic $\tan\beta$ dependence discussed
in Sec.~\ref{sec:one}, which is exact for $D=A$, but only approximate for
$D=h$.
For $m_h\alt120$~GeV, the $m_h$ dependence is approximately exponential,
following the heuristic law
$\sigma(pp\to Ch+X)\propto\exp[-4(m_h/100~\mbox{GeV})]$.
On the other hand, the $m_A$ dependence is approximately power-like over the
range of values considered, a heuristic law being
$\sigma(pp\to CA+X)\propto m_A^{-5}$.

We now move on to the predictions for Run~II at the Tevatron.
Comparing Figs.~\ref{fig:t}(a)--(c) with Figs.~\ref{fig:l}(a)--(c), we observe
that all distributions are approximately scaled down by a factor of 50 as we
pass from the LHC to the Tevatron.
Apart from that, their qualitative features essentially remain the same.

\begin{table}[ht]
\begin{center}
\caption{Total cross sections $\sigma$ in fb of $pp\to CD+X$ at the LHC and of
$p\overline{p}\to CD+X$ in Run~II at the Tevatron, where
$C=\Upsilon(1S),\chi_{bJ}(1P)$ and $D=h,A$, for $\tan\beta=50$ and
$m_D=100$~GeV.
The values enclosed in parentheses refer to the CSM.}
\label{tab:xs}
\medskip
\begin{tabular}{|c|cc|}
\hline\hline
$C+D$ & LHC & Tevatron \\
\hline
$\Upsilon+h$ & $3.3\times10^2$ & $6.0$ \\
$\chi_{bJ}+h$ & $7.4\times10^2$ ($1.9\times10^2$) & $1.3\times10^1$ (3.6) \\
$\Upsilon+A$ & $3.4\times10^2$ & $6.1$ \\
$\chi_{bJ}+A$ & $7.4\times10^2$ ($1.9\times10^2$) & $1.3\times10^1$ (3.6) \\
\hline\hline
\end{tabular}
\end{center}
\end{table}

We now assess the observability of the various processes at the LHC and in
Run~II the Tevatron.
To this end, we list their total cross sections for our MSSM benchmark
scenario in Table~\ref{tab:xs}.
These values can be converted into annual yields by recalling from Table~I of
Ref.~\cite{jb} that a cross section of 0.005~fb (0.25~fb) corresponds to one
event per year at the LHC (Tevatron).
At hadron colliders, the $\Upsilon(1S)$ mesons are easily detected through
their decays to $e^+e^-$ and $\mu^+\mu^-$ pairs, with a combined branching
fraction of $B(\Upsilon(1S)\to l^+l^-)=4.86\pm0.13$ \cite{pdg}.
The $\chi_{bJ}(1P)$ mesons radiatively decay to $\Upsilon(1S)$ mesons, the
individual branching fractions being
$B(\chi_{b0}(1P)\to\Upsilon(1S)\gamma)<6\%$,
$B(\chi_{b1}(1P)\to\Upsilon(1S)\gamma)=(35\pm8)\%$, and
$B(\chi_{b2}(1P)\to\Upsilon(1S)\gamma)=(22\pm4)\%$ \cite{pdg}.
In the high-$\tan\beta$ regime, light $h$ and $A$ bosons predominantly decay
to $b\overline{b}$ and $\tau^+\tau^-$ pairs, with branching fractions of about
90\% and 8\%, respectively \cite{pdg}.
The $b$ hadrons can be detected by looking for displaced decay vertices with
dedicated vertex detectors, even at the LHC.
The corresponding numbers of signal events per year estimated assuming
detection efficiencies of 100\% are listed in Table~\ref{tab:num}.

\begin{table}[ht]
\begin{center}
\caption{Annual numbers of events of $pp\to CD+X$ at the LHC and of
$p\overline{p}\to CD+X$ in Run~II at the Tevatron, where
$C=\Upsilon(1S),\chi_{bJ}(1P)$ and $D=h,A$, for $\tan\beta=50$ and
$m_D=100$~GeV.
The final-state particles are assumed to be detected via their decays
$\chi_{bJ}(1P)\to\Upsilon(1S)\gamma$, $\Upsilon(1S)\to e^+e^-,\mu^+\mu^-$, and
$h,A\to b\overline{b},\tau^+\tau^-$ with efficiencies of 100\%.
The values enclosed in parentheses refer to the CSM.}
\label{tab:num}
\medskip
\begin{tabular}{|c|cc|}
\hline\hline
$C+D$ & LHC & Tevatron \\
\hline
$\Upsilon+h$ & $3.1\times10^3$ & $1.1$ \\
$\chi_{bJ}+h$ & $1.7\times10^3$ ($4.4\times10^2$) & $6.2\times10^{-1}$
($1.6\times10^{-1}$) \\
$\Upsilon+A$ & $3.2\times10^3$ & $1.2$ \\
$\chi_{bJ}+A$ & $1.7\times10^3$ ($4.5\times10^2$) & $6.2\times10^{-1}$
($1.6\times10^{-1}$) \\
\hline\hline
\end{tabular}
\end{center}
\end{table}

We conclude this section by assessing the theoretical uncertainties arising
from the lack of knowledge of the precise values of the bottomonium MEs and
from the freedom in the choice of the renormalization and factorization scales.
Besides the QCD-motivated potential \cite{buc}, the authors of Ref.~\cite{eic}
also employed three other phenomenological ans\"atze, namely, a
power-law potential \cite{mar}, a logarithmic potential \cite{qui}, and a
Coulomb-plus-linear potential {\it(Cornell potential)} \cite{got}.
Taking as the theoretical errors the maximum upward and downward deviations
with respect to the results obtained with the QCD-motivated potential, we have
$\left\langle{\cal O}^{\Upsilon(1S)}\left[{}^3\!S_1^{(1)}\right]\right\rangle
=9.28{+10.85\atop-2.70}$~GeV${}^3$ and
$\left\langle{\cal O}^{\chi_{b0}(1P)}\left[{}^3\!P_0^{(1)}\right]\right\rangle
=2.03{+0.93\atop-0.00}$~GeV${}^5$.
Since there are no errors assigned to the CO MEs quoted in Ref.~\cite{kra},
for the purpose of this error analysis, we adopt the older results from
Ref.~\cite{cho2}, which are fitted to earlier CDF data \cite{fab}.
In compliance with Ref.~\cite{kra} (see also the discussion in the context of
Eq.~(4.1) in Ref.~\cite{cho2}), we assume that
$\left\langle{\cal O}^{\Upsilon(1S)}\left[{}^3\!P_0^{(8)}\right]\right\rangle
=m_b^2
\left\langle{\cal O}^{\Upsilon(1S)}\left[{}^1\!S_0^{(8)}\right]\right\rangle$.
We then repeat the evaluation of Table~\ref{tab:xs} with this set of
bottomonium MEs, varying one ME at a time and combining the individual shifts
in quadrature thereby allowing for the upper and lower half-errors to be
different.
The outcome is presented in Table~\ref{tab:xs1}. 
From Table~\ref{tab:xs1}, we observe that the NRQCD results for
$C=\Upsilon(1S),\chi_{bJ}(1P)$ carry errors of $\pm84\%$ and $\pm29\%$,
respectively, while the CSM result for $C=\chi_{bJ}(1P)$ may be increased by
up to 47\%.
We note in passing that CO MEs of Ref.~\cite{kra} fall outside the error bars
of their counterparts of Ref.~\cite{cho2}, so that the NRQCD results given in
Tables~\ref{tab:xs} and \ref{tab:xs1} are not actually mutually consistent.
While the face values of the NRQCD results in Table~\ref{tab:xs1} might by now
be obsolete, we believe that the relative errors should still serve as a
useful indicator.
The CDF data \cite{fab} from which the CO MEs of Ref.~\cite{cho2} were
extracted are only based on an integrated luminosity of 16.6~pb${}^{-1}$
collected in 1992--1993.
The design value of the integrated luminosity to be delivered by the Tevatron
until the end of 2009 is currently quoted as 8~fb${}^{-1}$ \cite{wit}.
The combined data sample of the CDF and D0 collaborations would then be three
orders of magnitude larger than the one used in Ref.~\cite{fab}, leading to a
reduction in statistical error by a factor of about 30.
The experimental errors in the cross section measurements of bottomonium
hadroproduction at the Tevatron should then be dominated by systematics.

\begin{table}[ht]
\begin{center}
\caption{Same as in Table~\ref{tab:xs}, but using the CO MEs of
Ref.~\cite{cho2} and taking into account the theoretical errors on the CS
\cite{eic} and CO \cite{cho2} MEs.}
\label{tab:xs1}
\medskip
\begin{tabular}{|c|cc|}
\hline\hline
$C+D$ & LHC & Tevatron \\
\hline
$\Upsilon+h$ & $(4.4\pm3.7)\times10^1$ & $(8.1\pm6.8)\times10^{-1}$ \\
$\chi_{bJ}+h$ & $(2.1\pm0.6)\times10^3$
$\left(1.9{+0.9\atop-0.0}\times10^2\right)$
& $(3.8\pm1.1)\times10^1$ $\left(3.6{+1.7\atop-0.0}\right)$ \\
$\Upsilon+A$ & $(4.4\pm3.7)\times10^1$ & $(8.2\pm6.8)\times10^{-1}$ \\
$\chi_{bJ}+A$ & $(2.1\pm0.6)\times10^3$
$\left(1.9{+0.9\atop-0.0}\times10^2\right)$
& $(3.8\pm1.1)\times10^1$ $\left(3.6{+1.6\atop-0.0}\right)$ \\
\hline\hline
\end{tabular}
\end{center}
\end{table}

Another plausible choice of renormalization and factorization scales is
$\mu_r=\mu_f=m_D$.
For $p_T\ll m_D$ ($p_T\gg m_D$), this scale choice is larger (smaller) than
our default setting.
The $p_T$ distributions are particularly sensitive to this change.
In order to illustrate this, we recalculated the $p_T$ distributions shown in
the upper panels of Figs.~\ref{fig:l}(a) and \ref{fig:t}(a) using this
alternative scale choice.
The results are displayed in the upper and lower panels of Fig.~\ref{fig:s},
respectively.
From Fig.~\ref{fig:s}, we observe that the $p_T$ distributions are appreciably
reduced in the lower $p_T$ range, for $p_T\alt25$~GeV, as we pass from our
default scale choice to the new one.
This feature is more pronounced for the Tevatron than for the LHC.
On the other hand, towards the upper end of the $p_T$ values considered in the
case of the LHC, the shifts in cross section induced by switching to the new
scale choice are insignificant.
Such scale uncertainties are generally expected to be reduced as
next-to-leading-order corrections are taken into account.

\section{Conclusions}
\label{sec:four}

We studied the associated production of heavy quarkonia $C$ and Higgs bosons
$D$ in photon-photon, photon-hadron, and hadron-hadron collisions to LO in the
NRQCD factorization formalism.
We considered all experimentally established heavy quarkonia, with
${}^{2S+1}L_J={}^1\!S_0,{}^3\!S_1,{}^1\!P_1,{}^3\!P_J$ ($J=0,1,2$), and the SM
Higgs boson $D={\cal H}$ as well as the MSSM Higgs bosons $D=h,H,A,H^\pm$.
We listed the cross sections of all contributing partonic subprocesses, except
for those with $D=H^\pm$, which are either suppressed by small Yukawa
couplings or involve an additional heavy particle with a distinct signature in
the final state.
We presented numerical results for any combination of
$C=\Upsilon(1S),\chi_{bJ}(1P)$ and $D=h,A$ appropriate for the LHC and Run~II
at the Tevatron, concentrating on the region of the currently allowed MSSM
parameter space characterized by a large value of $\tan\beta$ and small values
of $m_h$ and $m_A$.

Criteria to discriminate NRQCD against the CSM include the following:
(i) $\Upsilon+h$ and $\Upsilon+A$ associated production is allowed in
NRQCD, but forbidden to LO in the CSM; and
(ii) in NRQCD, the cross sections of $\chi_{bJ}+h$ and $\chi_{bJ}+A$
associated production are a factor of 3 to 4 larger than in the CSM.

In our MSSM benchmark scenario, there will be about 3000 (1) direct
$\Upsilon(1S)+h\to l^+l^-+b\overline{b}$ events per year at the LHC (Tevatron)
and similarly for the $A$ boson.
The $\chi_{bJ}\to\Upsilon(1S)\gamma$ feed-down channels will add about 1500
(0.5) events to each of these yields.
Results in the same ballpark are expected for the bottomonia with $n=2,3$.
We conclude that these signals should be clearly visible at the LHC, while
they should provide a challenge for Run~II at the Tevatron.

\bigskip

\centerline{\bf Acknowledgments}

\smallskip

B.A.K. and C.P.P. are grateful to the Kavli Institute for Theoretical Physics
at the University of California at Santa Barbara and the Second Institute for
Theoretical Physics at the University of Hamburg, respectively, for their
hospitality during visits when this paper was prepared.
The research of B.A.K. was supported in part by the Deutsche
Forschungsgemeinschaft (DFG) under Grant No.\ KN~365/1, by the
Bundesministerium f\"ur Bildung und Forschung under Grant No.\ 05~HT1GUA/4,
and by the National Science Foundation under Grant No.\ PHY99-07949. 
The research of C.P.P. was supported in part by the DFG through
Graduiertenkolleg No.\ GRK~602/1 and by the Office of the Vice President for
Academic Affairs of the University of the Philippines.
The research of L.Z. was supported by the Studienstiftung des deutschen Volkes
through a PhD scholarship.

\newpage

\def\theequation{\Alph{section}.\arabic{equation}}
\begin{appendix}
\setcounter{equation}{0}
\section{Partonic cross sections}

In this appendix, we list the differential cross sections $d\sigma/dt$ for
processes (\ref{eq:gg})--(\ref{eq:pp}) with $D=h,A$.
The results for $D={\cal H},H$ may be obtained from those for $D=h$ by
accordingly replacing $y_Q^h$ and $m_h$.
Expressions of the partonic Mandelstam variables $s$, $t$, and $u$ in terms
of $p_T$, $y_C$, and $y_D$ may be found in Eq.~(4) of Ref.~\cite{jb}.
The mass $m_C=M$ of the heavy quarkonium is taken to be $M=2m_Q$.

\begin{eqnarray}
&&\frac{d\sigma}{dt}\left(g g \to Q\overline{Q}\left[{}^1\!
S_0^{(1)}\right] h\right)
=\frac{2 \pi \alpha_s^2 \left(y_Q^h\right)^2 M (2 m_h^2 - s - t - u)^2}
{9 (m_h^2 - s - t)^2 (m_h^2 - s - u)^2 (2 m_h^2 - t - u)^2},
\\
&&\frac{d\sigma}{dt}\left(g g \to Q\overline{Q}\left[{}^3\!
P_J^{(1)}\right] h\right)
=\frac{-4 \pi \alpha_s^2 \left(y_Q^h\right)^2}{135 M^3 s^2 (m_h^2 - s - t)^4
(m_h^2 - s - u)^4 (2 m_h^2 - t - u)^4}
\nonumber\\
&&{}\times F_J,\qquad J=0,1,2,
\\
&&F_0
= -10 M^2 \{ 196 m_h^{20} - 
   140 m_h^{18} [8 s + 7 (t + u)]
\nonumber\\
&&{} + m_h^{16} [2756 s^2 + 4844 s (t + u) + 
     35 (59 t^2 + 134 t u + 59 u^2)]
\nonumber\\
&&{} - 
   2 m_h^{14} [1920 s^3 + 5092 s^2 (t + u) + 
     s (4329 t^2 + 9934 t u + 4329 u^2)
\nonumber\\
&&{} + 
     70 (t + u) (17 t^2 + 50 t u + 17 u^2)]
\nonumber\\
&&{} + 
   2 m_h^{12} [1680 s^4 + 6012 s^3 (t + u) 
+ 4 s^2 (1908 t^2 + 4409 t u + 1908 u^2)
\nonumber\\
&&{} + 
     7 s (t + u) (589 t^2 + 1790 t u + 589 u^2)
\nonumber\\
&&{} + 2 (407 t^4 + 2537 t^3 u + 4402 t^2 u^2 + 2537 t u^3 + 407 u^4)]
\nonumber\\
&&{} - 
   2 m_h^{10} [968 s^5 + 4462 s^4 (t + u) + 
     s^3 (7559 t^2 + 17434 t u + 7559 u^2)
\nonumber\\
&&{} + 
     2 s^2 (t + u) (2983 t^2 + 9210 t u + 2983 u^2)
\nonumber\\
&&{} + 
     s (2245 t^4 + 14659 t^3 u + 25776 t^2 u^2 + 14659 t u^3 + 2245 u^4)
\nonumber\\
&&{} + (t + u) (335 t^4 + 2874 t^3 u + 5930 t^2 u^2 + 2874 t u^3 + 335 u^4)]
\nonumber\\
&&{} + 
   m_h^8 [740 s^6 + 4412 s^5 (t + u) + 2 s^4 (4803 t^2 + 10882 t u + 4803 u^2)
\nonumber\\
&&{} + 
     2 s^3 (t + u) (4959 t^2 + 14882 t u + 4959 u^2)
\nonumber\\
&&{} + 
     s^2 (5219 t^4 + 34390 t^3 u + 60782 t^2 u^2 + 34390 t u^3 + 5219 u^4)
\nonumber\\
&&{} + 
     14 s (t + u) (99 t^4 + 934 t^3 u + 1994 t^2 u^2 + 934 t u^3 + 99 u^4)
\nonumber\\
&&{} + 10 (16 t^6 + 239 t^5 u + 1007 t^4 u^2 + 1592 t^3 u^3 + 
       1007 t^2 u^4 + 239 t u^5 + 16 u^6)]
\nonumber\\
&&{} - 
   2 m_h^6 [88 s^7 + 728 s^6 (t + u) + s^5 (2089 t^2 + 4574 t u + 2089 u^2)
\nonumber\\
&&{} + 
     26 s^4 (t + u) (107 t^2 + 292 t u + 107 u^2)
\nonumber\\
&&{} + 
     5 s^3 (375 t^4 + 2270 t^3 u + 3942 t^2 u^2 + 2270 t u^3 + 375 u^4)
\nonumber\\
&&{} + 
     s^2 (t + u) (643 t^4 + 5906 t^3 u + 12746 t^2 u^2 + 5906 t u^3 + 643 u^4)
\nonumber\\
&&{} + s (111 t^6 + 1965 t^5 u + 8844 t^4 u^2 + 14224 t^3 u^3 + 
       8844 t^2 u^4 + 1965 t u^5 + 111 u^6)
\nonumber\\
&&{} + 
     10 (t + u) (t^2 + 5 t u + u^2) (t^4 + 19 t^3 u + 44 t^2 u^2 + 19 t u^3 
+ u^4)]
\nonumber\\
&&{} + 
   m_h^4 [20 s^8 + 
     288 s^7 (t + u) + 8 s^6 (149 t^2 + 314 t u + 149 u^2)
\nonumber\\
&&{} + 
     2 s^5 (t + u) (1077 t^2 + 2578 t u + 1077 u^2)
\nonumber\\
&&{} + 
     2 s^4 (969 t^4 + 4952 t^3 u + 8206 t^2 u^2 + 4952 t u^3 + 969 u^4)
\nonumber\\
&&{} + 
     2 s^3 (t + u) (437 t^4 + 3048 t^3 u + 6210 t^2 u^2 + 3048 t u^3 + 437 u^4)
\nonumber\\
&&{} + s^2 (179 t^6 + 2630 t^5 u + 11681 t^4 u^2 + 18788 t^3 u^3 + 
       11681 t^2 u^4 + 2630 t u^5 + 179 u^6)
\nonumber\\
&&{} + 
     2 s (t + u) (7 t^6 + 258 t^5 u + 1740 t^4 u^2 + 3326 t^3 u^3 
+ 1740 t^2 u^4 + 258 t u^5 + 7 u^6)
\nonumber\\
&&{} + t^8 + 52 t^7 u + 568 t^6 u^2 + 2144 t^5 u^3 + 
     3290 t^4 u^4 + 2144 t^3 u^5 + 568 t^2 u^6 + 52 t u^7 + u^8]
\nonumber\\
&&{} - 
   2 m_h^2 [12 s^8 (t + u) + s^7 (91 t^2 + 186 t u + 91 u^2)
%\nonumber\\
%&&{}
+ 2 s^6 (t + u) (121 t^2 + 260 t u + 121 u^2)
\nonumber\\
&&{} + 
     s^5 (308 t^4 + 1361 t^3 u + 2142 t^2 u^2 + 1361 t u^3 + 308 u^4)
\nonumber\\
&&{} + 
     2 s^4 (t + u) (101 t^4 + 511 t^3 u + 919 t^2 u^2 + 511 t u^3 + 101 u^4)
\nonumber\\
&&{} + 
     s^3 (65 t^6 + 566 t^5 u + 2070 t^4 u^2 + 3182 t^3 u^3 + 2070 t^2 u^4 + 
       566 t u^5 + 65 u^6)
\nonumber\\
&&{} + 
     s^2 (t + u) (8 t^6 + 119 t^5 u + 744 t^4 u^2 + 1418 t^3 u^3 + 744 t^2 u^4
+ 119 t u^5 + 8 u^6)
\nonumber\\
&&{} + s t u (13 t^6 + 204 t^5 u + 852 t^4 u^2 + 1334 t^3 u^3 + 
       852 t^2 u^4 + 204 t u^5 + 13 u^6)
\nonumber\\
&&{} + t u (t + u) (t^2 + 5 t u + u^2) (t^4 + 16 t^3 u + 36 t^2 u^2 + 16 t u^3
+ u^4)]
\nonumber\\
&&{} + 9 s^8 (t + u)^2 + 42 s^7 (t + u)^3 + 
   s^6 (79 t^4 + 322 t^3 u + 490 t^2 u^2 + 322 t u^3 + 79 u^4)
\nonumber\\
&&{} + 4 s^5 (t + u) (19 t^4 + 81 t^3 u + 131 t^2 u^2 + 81 t u^3 + 19 u^4)
\nonumber\\
&&{} + s^4 (39 t^6 + 258 t^5 u + 755 t^4 u^2 + 
     1080 t^3 u^3 + 755 t^2 u^4 + 258 t u^5 + 39 u^6)
\nonumber\\
&&{} + 
   2 s^3 (t + u) (5 t^6 + 36 t^5 u + 144 t^4 u^2 + 246 t^3 u^3 + 144 t^2 u^4 + 
  36 t u^5 + 5 u^6)
\nonumber\\
&&{} + 
   s^2 (t^8 + 10 t^7 u + 91 t^6 u^2 + 356 t^5 u^3 + 552 t^4 u^4 + 
     356 t^3 u^5 + 91 t^2 u^6 + 10 t u^7 + u^8)
\nonumber\\
&&{} + 
   4 s t^2 u^2 (t + u) (t^2 + 5 t u + u^2) (3 t^2 + 7 t u + 3 u^2)
\nonumber\\
&&{} + 
   t^2 u^2 (t + u)^2 (t^2 + 5 t u + u^2)^2\} ,  
\\
&&F_1
= -5 (2 m_h^2 - s - t - u)^2\{ 4 m_h^{16} s - 
   4 m_h^{14} [10 s^2 + 4 s (t + u) + (t - u)^2]
\nonumber\\
&&{} + 
   2 m_h^{12} [62 s^3 + 72 s^2 (t + u) + 
     s (23 t^2 + 10 t u + 23 u^2) + 8 (t + u) (t - u)^2]
\nonumber\\
&&{} - m_h^{10} [176 s^4 + 368 s^3 (t + u) + 
     2 s^2 (119 t^2 + 206 t u + 119 u^2)
\nonumber\\
&&{} + 2 s (t + u) (47 t^2 - 38 t u + 47 u^2) + 
     (t - u)^2 (25 t^2 + 58 t u + 25 u^2)]
\nonumber\\
&&{} + m_h^8 [124 s^5 + 400 s^4 (t + u) + 
     4 s^3 (109 t^2 + 234 t u + 109 u^2)
\nonumber\\
&&{} + 4 s^2 (t + u) (61 t^2 + 68 t u + 61 u^2) + 
     s (113 t^4 + 86 t^3 u - 118 t^2 u^2 + 86 t u^3 + 113 u^4)
\nonumber\\
&&{} + (t + u) (t - u)^2 (19 t^2 + 62 t u + 19 u^2)]
\nonumber\\
&&{} - 
   m_h^6 [40 s^6 + 192 s^5 (t + u) + 8 s^4 (41 t^2 + 96 t u + 41 u^2)
\nonumber\\
&&{} + 
     8 s^3 (t + u) (34 t^2 + 79 t u + 34 u^2)
\nonumber\\
&&{} + 
     2 s^2 (83 t^4 + 194 t^3 u + 226 t^2 u^2 + 194 t u^3 + 83 u^4)
\nonumber\\
&&{} + 
     4 s (t + u) (19 t^4 + 17 t^3 u - 44 t^2 u^2 + 17 t u^3 + 19 u^4)
\nonumber\\
&&{} + 
     (t - u)^2 (7 t^4 + 54 t^3 u + 98 t^2 u^2 + 54 t u^3 + 7 u^4)]
\nonumber\\
&&{} + 
   m_h^4 [4 s^7 + 32 s^6 (t + u) + 
     2 s^5 (47 t^2 + 114 t u + 47 u^2)
\nonumber\\
&&{} + 4 s^4 (t + u) (31 t^2 + 92 t u + 31 u^2)
\nonumber\\
&&{} + 2 s^3 (51 t^4 + 206 t^3 u + 340 t^2 u^2 + 
       206 t u^3 + 51 u^4) + 10 s^2 (t + u)^3 (7 t^2 - 2 t u + 7 u^2)
\nonumber\\
&&{} + s (27 t^6 + 100 t^5 u - t^4 u^2 - 
       140 t^3 u^3 - t^2 u^4 + 100 t u^5 + 27 u^6)
\nonumber\\
&&{} + (t + u) (t - u)^2 (t^4 + 16 t^3 u + 38 t^2 u^2 + 16 t u^3 + u^4)]
\nonumber\\
&&{} - 
   m_h^2 [6 s^6 (t + u)^2 + 
     2 s^5 (t + u) (9 t^2 + 26 t u + 9 u^2)
\nonumber\\
&&{} + 
     s^4 (25 t^4 + 124 t^3 u + 214 t^2 u^2 + 124 t u^3 + 25 u^4)
\nonumber\\
&&{} + 
     4 s^3 (t + u) (6 t^4 + 17 t^3 u + 36 t^2 u^2 + 17 t u^3 + 6 u^4)
\nonumber\\
&&{} + 
     s^2 (15 t^6 + 56 t^5 u + 63 t^4 u^2 + 60 t^3 u^3 + 63 t^2 u^4 + 
       56 t u^5 + 15 u^6)
\nonumber\\
&&{} + 2 s (t + u) (2 t^6 + 14 t^5 u - 3 t^4 u^2 - 18 t^3 u^3 - 3 t^2 u^4 + 
  14 t u^5 + 2 u^6)
\nonumber\\
&&{} + t u (t^2 - u^2)^2 (2 t^2 + 9 t u + 2 u^2)]
\nonumber\\
&&{} + 
   s^5 (t^4 + 6 t^3 u + 6 t^2 u^2 + 6 t u^3 + u^4)
\nonumber\\
&&{} + 
   s^4 (t + u) (3 t^4 + 8 t^3 u + 18 t^2 u^2 + 8 t u^3 + 3 u^4)
\nonumber\\
&&{} + 
   s^3 (3 t^6 + 8 t^5 u + 25 t^4 u^2 + 32 t^3 u^3 + 25 t^2 u^4 + 8 t u^5 + 
     3 u^6)
\nonumber\\
&&{} + s^2 (t + u) (t^6 + 6 t^5 u + 3 t^4 u^2 + 4 t^3 u^3 + 3 t^2 u^4 
+ 6 t u^5 + u^6)
\nonumber\\
&&{} + s t u (4 t^6 + 5 t^5 u - 2 t^4 u^2 - 10 t^3 u^3 - 2 t^2 u^4 + 
     5 t u^5 + 4 u^6)
\nonumber\\
&&{} + t^2 u^2 (t - u)^2 (t + u)^3\} ,  
\\
&&F_2
= 448 m_h^{22} - 
   16 m_h^{20} [149 s + 168 (t + u)]
\nonumber\\
&&{} + 16 m_h^{18} [354 s^2 + 800 s (t + u) + 
     7 (63 t^2 + 134 t u + 63 u^2)]
\nonumber\\
&&{} - 4 m_h^{16} [2111 s^3 + 6834 s^2 (t + u) + 
     s (7491 t^2 + 15766 t u + 7491 u^2)
\nonumber\\
&&{} + 
     140 (t + u) (19 t^2 + 46 t u + 19 u^2)]
\nonumber\\
&&{} + 
   4 m_h^{14} [2358 s^4 + 9272 s^3 (t + u) + 
     s^2 (14369 t^2 + 29782 t u + 14369 u^2)
\nonumber\\
&&{} + 
     s (t + u) (10041 t^2 + 23470 t u + 10041 u^2)
\nonumber\\
&&{} + 
     7 (363 t^4 + 1796 t^3 u + 2882 t^2 u^2 + 1796 t u^3 + 363 u^4)]
\nonumber\\
&&{} - 
   2 m_h^{12} [4130 s^5 + 18820 s^4 (t + u) + 
     5 s^3 (7093 t^2 + 14430 t u + 7093 u^2)
\nonumber\\
&&{} + 
     26 s^2 (t + u) (1329 t^2 + 2954 t u + 1329 u^2)
\nonumber\\
&&{} + 
     s (16995 t^4 + 80520 t^3 u + 127418 t^2 u^2 + 80520 t u^3 + 16995 u^4)
\nonumber\\
&&{} + 
     56 (t + u) (57 t^4 + 332 t^3 u + 566 t^2 u^2 + 332 t u^3 + 57 u^4)]
\nonumber\\
&&{} + m_h^{10} [5232 s^6 + 28832 s^5 (t + u) + 
     2 s^4 (32427 t^2 + 64870 t u + 32427 u^2)
\nonumber\\
&&{} + 
     2 s^3 (t + u) (38653 t^2 + 80838 t u + 38653 u^2)
\nonumber\\
&&{} + s^2 (52313 t^4 + 233504 t^3 u + 362542 t^2 u^2 + 
       233504 t u^3 + 52313 u^4)
\nonumber\\
&&{} + 4 s (t + u) (4710 t^4 + 25333 t^3 u + 41834 t^2 u^2 + 25333 t u^3 
+ 4710 u^4)
\nonumber\\
&&{} + 
     7 (t + u)^2 (377 t^4 + 2748 t^3 u + 5174 t^2 u^2 + 2748 t u^3 + 
       377 u^4)]
\nonumber\\
&&{} - 
   m_h^8 [2116 s^7 + 15080 s^6 (t + u) + 224 s^5 (191 t^2 + 377 t u + 
       191 u^2)
\nonumber\\
&&{} + 4 s^4 (t + u) (15689 t^2 + 31000 t u + 15689 u^2)
\nonumber\\
&&{} + 
     s^3 (52571 t^4 + 218938 t^3 u + 332542 t^2 u^2 + 218938 t u^3 + 
       52571 u^4)
\nonumber\\
&&{} + s^2 (t + u) (25783 t^4 + 124384 t^3 u + 197458 t^2 u^2 + 124384 t u^3 + 
  25783 u^4)
\nonumber\\
&&{} + 
     s (6841 t^6 + 57484 t^5 u + 171535 t^4 u^2 + 241720 t^3 u^3
\nonumber\\
&&{} + 
       171535 t^2 u^4 + 57484 t u^5 + 6841 u^6)
\nonumber\\
&&{} + 
     7 (t + u)^3 (99 t^4 + 980 t^3 u + 2162 t^2 u^2 + 980 t u^3 + 99 u^4)]
\nonumber\\
&&{} + 
   m_h^6 [472 s^8 + 4752 s^7 (t + u) + 8 s^6 (2279 t^2 + 4436 t u + 
       2279 u^2)
\nonumber\\
&&{} + 4 s^5 (t + u) (8689 t^2 + 16336 t u + 8689 u^2)
\nonumber\\
&&{} + 
     2 s^4 (18433 t^4 + 72142 t^3 u + 107398 t^2 u^2 + 72142 t u^3 + 
       18433 u^4)
\nonumber\\
&&{} + 4 s^3 (t + u) (5706 t^4 + 24309 t^3 u + 37018 t^2 u^2 + 24309 t u^3 + 
  5706 u^4)
\nonumber\\
&&{} + 
     s^2 (8227 t^6 + 60592 t^5 u + 168433 t^4 u^2 + 232088 t^3 u^3
\nonumber\\
&&{} + 
       168433 t^2 u^4 + 60592 t u^5 + 8227 u^6)
\nonumber\\
&&{} + 
     2 s (t + u) (785 t^6 + 7956 t^5 u + 25901 t^4 u^2 + 37404 t^3 u^3
\nonumber\\
&&{}  + 
  25901 t^2 u^4 + 7956 t u^5 + 785 u^6)
\nonumber\\
&&{} + 
     35 (t + u)^4 (3 t^4 + 46 t^3 u + 130 t^2 u^2 + 46 t u^3 + 3 u^4)]
\nonumber\\
&&{} - 
   m_h^4 [44 s^9 + 776 s^8 (t + u) + 2 s^7 (2229 t^2 + 4262 t u + 2229 u^2)
\nonumber\\
&&{} + 
     8 s^6 (t + u) (1453 t^2 + 2593 t u + 1453 u^2)
\nonumber\\
&&{} + 
     4 s^5 (4066 t^4 + 15109 t^3 u + 22095 t^2 u^2 + 15109 t u^3 + 
       4066 u^4)
\nonumber\\
&&{} + 2 s^4 (t + u) (6549 t^4 + 25090 t^3 u + 36846 t^2 u^2 + 25090 t u^3 + 
  6549 u^4)
\nonumber\\
&&{} + 
     s^3 (6151 t^6 + 39452 t^5 u + 101439 t^4 u^2 + 136332 t^3 u^3
\nonumber\\
&&{} + 
       101439 t^2 u^4 + 39452 t u^5 + 6151 u^6)
\nonumber\\
&&{} + 
     s^2 (t + u) (1619 t^6 + 13326 t^5 u + 38633 t^4 u^2 + 53884 t^3 u^3
\nonumber\\
&&{}  + 
  38633 t^2 u^4 + 13326 t u^5 + 1619 u^6)
\nonumber\\
&&{} + 
     s (207 t^8 + 3214 t^7 u + 16050 t^6 u^2 + 38562 t^5 u^3 + 
       50990 t^4 u^4
\nonumber\\
&&{} + 38562 t^3 u^5 + 16050 t^2 u^6 + 3214 t u^7 + 
       207 u^8)
\nonumber\\
&&{} + 
     7 (t + u)^5 (t^4 + 32 t^3 u + 134 t^2 u^2 + 32 t u^3 + u^4)]
\nonumber\\
&&{} + m_h^2 [48 s^9 (t + u) + 2 s^8 (271 t^2 + 510 t u + 271 u^2)
\nonumber\\
&&{} + 
     2 s^7 (t + u) (1023 t^2 + 1726 t u + 1023 u^2)
\nonumber\\
&&{} + 
     s^6 (3849 t^4 + 13644 t^3 u + 19606 t^2 u^2 + 13644 t u^3 + 3849 u^4)
\nonumber\\
&&{} + 
     4 s^5 (t + u) (1027 t^4 + 3626 t^3 u + 5160 t^2 u^2 + 3626 t u^3 
+ 1027 u^4)
\nonumber\\
&&{} + 2 s^4 (1290 t^6 + 7467 t^5 u + 
       17981 t^4 u^2 + 23632 t^3 u^3
%\nonumber\\
%&&{} 
+ 17981 t^2 u^4 + 7467 t u^5 + 
       1290 u^6)
\nonumber\\
&&{} + 10 s^3 (t + u) (93 t^6 + 626 t^5 u + 1608 t^4 u^2 + 2162 t^3 u^3 + 
  1608 t^2 u^4 + 626 t u^5 + 93 u^6)
\nonumber\\
&&{} + 
     s^2 (173 t^8 + 2052 t^7 u + 8737 t^6 u^2 + 19348 t^5 u^3 + 
       24948 t^4 u^4
\nonumber\\
&&{} + 19348 t^3 u^5 + 8737 t^2 u^6 + 2052 t u^7 + 173 u^8)
\nonumber\\
&&{} + 
     4 s (t + u) (3 t^8 + 76 t^7 u + 441 t^6 u^2 + 1135 t^5 u^3 + 1522 t^4 u^4
\nonumber\\
&&{}  + 
  1135 t^3 u^5 + 441 t^2 u^6 + 76 t u^7 + 3 u^8)
%\nonumber\\
%&&{} 
+ 7 t u (t + u)^6 
      (2 t^2 + 17 t u + 2 u^2)]
\nonumber\\
&&{} - 24 s^9 (t + u)^2 - 48 s^8 (t + u) (3 t^2 + 5 t u + 3 u^2)
\nonumber\\
&&{} - 
   s^7 (367 t^4 + 1258 t^3 u + 1786 t^2 u^2 + 1258 t u^3 + 367 u^4)
\nonumber\\
&&{} - 
   s^6 (t + u) (515 t^4 + 1708 t^3 u + 2370 t^2 u^2 + 1708 t u^3 + 515 u^4)
\nonumber\\
&&{} - 
   2 s^5 (215 t^6 + 1162 t^5 u + 2665 t^4 u^2 + 3440 t^3 u^3 + 2665 t^2 u^4 + 
  1162 t u^5 + 215 u^6)
\nonumber\\
&&{} - 2 s^4 (t + u) (107 t^6 + 630 t^5 u + 1477 t^4 u^2 + 1920 t^3 u^3
%\nonumber\\
%&&{}
+ 1477 t^2 u^4 + 630 t u^5 + 107 u^6)
\nonumber\\
&&{} - 
   s^3 (59 t^8 + 558 t^7 u + 2051 t^6 u^2 + 4160 t^5 u^3 + 5220 t^4 u^4
\nonumber\\
&&{} + 
     4160 t^3 u^5 + 2051 t^2 u^6 + 558 t u^7 + 59 u^8)
\nonumber\\
&&{} - 
   s^2 (t + u) (7 t^8 + 112 t^7 u + 505 t^6 u^2 + 1152 t^5 u^3 + 1496 t^4 u^4
\nonumber\\
&&{}  + 
  1152 t^3 u^5 + 505 t^2 u^6 + 112 t u^7 + 7 u^8)
\nonumber\\
&&{} - s t u (t + u)^2 (12 t^6 + 85 t^5 u + 242 t^4 u^2 + 
     326 t^3 u^3 + 242 t^2 u^4 + 85 t u^5 + 12 u^6)
\nonumber\\
&&{} - 
   7 t^2 u^2 (t + u)^7 ,  
\\
&&\frac{d\sigma}{dt}\left(g g \to Q\overline{Q}\left[
{}^1\!S_0^{(8)}\right] h\right)
=\frac{15}{8}\,\frac{d\sigma}{dt}\left(g g \to Q\overline{Q}\left[
{}^1\!S_0^{(1)}\right] h\right),
\\
&&\frac{d\sigma}{dt}\left(g g \to Q\overline{Q}\left[{}^3\!
S_1^{(8)}\right] h\right)
=\frac{\pi \alpha_s^2 \left(y_Q^h\right)^2}
{4 M s^3 (m_h^2 - s - t)^2 (m_h^2 - s - u)^2 (2 m_h^2 - t - u)^2}
\nonumber\\
&&{}\times
\{9 m_h^{12} - 9 m_h^{10} [2 s + 3 (t + u)]
+ m_h^8 [10 s^2 + 40 s (t + u) + 3 (11 t^2 + 23 t u + 11 u^2)]
\nonumber\\
&&{} - m_h^6 [s^3 + 15 s^2 (t + u) + 4 s (9 t^2 + 17 t u + 9 u^2) + 
     3 (t + u) (7 t^2 + 16 t u + 7 u^2)]
\nonumber\\
&&{} + 
   m_h^4 [s^3 (t + u) + 
     2 s^2 (5 t^2 + 6 t u + 5 u^2) + 4 s (t + u) (4 t^2 + 7 t u + 4 u^2)
\nonumber\\
&&{} + 7 t^4 + 35 t^3 u + 51 t^2 u^2 + 35 t u^3 + 7 u^4]
\nonumber\\
&&{} - 
   m_h^2 [s^3 (t^2 - t u + u^2) + s^2 (t + u) (3 t^2 + t u + 3 u^2)
\nonumber\\
&&{} + 
     s (3 t^4 + 14 t^3 u + 16 t^2 u^2 + 14 t u^3 + 3 u^4) + (t + u) (t^2 + t u + u^2) (t^2 + 7 t u + u^2)]
\nonumber\\
&&{} + 
   t u [s^2 (t^2 + u^2) + 
     2 s (t + u) (t^2 + u^2) + (t^2 + t u + u^2)^2]
 \} ,
\\
&&\frac{d\sigma}{dt}\left(g g \to Q\overline{Q}\left[{}^1\!
P_1^{(8)}\right] h\right)
=\frac{\pi \alpha_s^2 \left(y_Q^h\right)^2}
{M^3 s^3 (m_h^2 - s - t)^3 (m_h^2 - s - u)^3 (2 m_h^2 - t - u)^4}
\nonumber\\
&&{}\times
\{ 36 m_h^{20} - 4 m_h^{18} [38 s + 45 (t +
u)]
\nonumber\\
&&{}  + 
   m_h^{16} [300 s^2 + 668 s (t + u) + 3 (131 t^2 + 278 t u + 131 u^2)]
\nonumber\\
&&{}  - 
   2 m_h^{14} [182 s^3 + 586 s^2 (t + u) + s (639 t^2 + 1330 t u + 639 u^2)
\nonumber\\
&&{}  + 
     6 (t + u) (41 t^2 + 98 t u + 41 u^2)]
\nonumber\\
&&{}  + 
   m_h^{12} [308 s^4 + 1256 s^3 (t + u) + s^2 (1979 t^2 + 4050 t u + 1979 u^2)
\nonumber\\
&&{}  + 
     7 s (t + u) (199 t^2 + 450 t u + 199 u^2) + 
     4 (97 t^4 + 473 t^3 u + 750 t^2 u^2 + 473 t u^3 + 97 u^4)]
\nonumber\\
&&{}  - m_h^{10} [204 s^5 + 940 s^4 (t + u) + 
     s^3 (1841 t^2 + 3722 t u + 1841 u^2)
\nonumber\\
&&{}  + 
     2 s^2 (t + u) (937 t^2 + 2032 t u + 937 u^2)
\nonumber\\
&&{}  + 
     7 s (135 t^4 + 622 t^3 u + 966 t^2 u^2 + 622 t u^3 + 135 u^4)
\nonumber\\
&&{}  + 
     6 (t + u) (33 t^4 + 190 t^3 u + 310 t^2 u^2 + 190 t u^3 + 33 u^4)]
\nonumber\\
&&{}  + m_h^8 [104 s^6 + 552 s^5 (t + u) + 
     2 s^4 (597 t^2 + 1198 t u + 597 u^2)
\nonumber\\
&&{}  + 
     4 s^3 (t + u) (367 t^2 + 776 t u + 367 u^2)
\nonumber\\
&&{}  + 
     s^2 (1074 t^4 + 4779 t^3 u + 7334 t^2 u^2 + 4779 t u^3 + 1074 u^4)
\nonumber\\
&&{}  + 
     s (t + u) (407 t^4 + 2145 t^3 u + 3352 t^2 u^2 + 2145 t u^3 + 407 u^4)
\nonumber\\
&&{}  + 2 (32 t^6 + 303 t^5 u + 915 t^4 u^2 + 1280 t^3 u^3 + 
       915 t^2 u^4 + 303 t u^5 + 32 u^6)]
\nonumber\\
&&{}  - 
   2 m_h^6 [16 s^7 + 116 s^6 (t + u) + 
     s^5 (297 t^2 + 602 t u + 297 u^2)
\nonumber\\
&&{}  + 
     2 s^4 (t + u) (197 t^2 + 418 t u + 197 u^2)
\nonumber\\
&&{}  + 
     s^3 (335 t^4 + 1481 t^3 u + 2258 t^2 u^2 + 1481 t u^3 + 335 u^4)
\nonumber\\
&&{}  + 
     2 s^2 (t + u) (93 t^4 + 473 t^3 u + 723 t^2 u^2 + 473 t u^3 + 93 u^4)
\nonumber\\
&&{}  + s (54 t^6 + 464 t^5 u + 1319 t^4 u^2 + 1814 t^3 u^3 + 
       1319 t^2 u^4 + 464 t u^5 + 54 u^6)
\nonumber\\
&&{}  + 2 (t + u) (3 t^6 + 40 t^5 u + 134 t^4 u^2 + 186 t^3 u^3 + 134 t^2 u^4 + 
  40 t u^5 + 3 u^6)]
\nonumber\\
&&{}  + 
   m_h^4 [4 s^8 + 52 s^7 (t + u) + 
     s^6 (189 t^2 + 394 t u + 189 u^2)
\nonumber\\
&&{}  + 
     3 s^5 (t + u) (101 t^2 + 230 t u + 101 u^2)
\nonumber\\
&&{}  + 
     2 s^4 (135 t^4 + 626 t^3 u + 960 t^2 u^2 + 626 t u^3 + 135 u^4)
\nonumber\\
&&{}  + 
     s^3 (t + u) (165 t^4 + 886 t^3 u + 1330 t^2 u^2 + 886 t u^3 + 165 u^4)
\nonumber\\
&&{}  + s^2 (72 t^6 + 623 t^5 u + 1733 t^4 u^2 + 2368 t^3 u^3 + 
       1733 t^2 u^4 + 623 t u^5 + 72 u^6)
\nonumber\\
&&{}  + 
     s (t + u) (16 t^6 + 185 t^5 u + 559 t^4 u^2 + 768 t^3 u^3 + 559 t^2 u^4 + 
  185 t u^5 + 16 u^6)
\nonumber\\
&&{}  + t^8 + 28 t^7 u + 160 t^6 u^2 + 376 t^5 u^3 + 490 t^4 u^4 + 
     376 t^3 u^5 + 160 t^2 u^6 + 28 t u^7 + u^8]
\nonumber\\
&&{}  - 
   m_h^2 [4 s^8 (t + u) + s^7 (27 t^2 + 58 t u + 27 u^2) + 
     2 s^6 (t + u) (31 t^2 + 80 t u + 31 u^2)
\nonumber\\
&&{}  + 
     s^5 (67 t^4 + 354 t^3 u + 562 t^2 u^2 + 354 t u^3 + 67 u^4)
\nonumber\\
&&{}  + 
     4 s^4 (t + u) (10 t^4 + 66 t^3 u + 101 t^2 u^2 + 66 t u^3 + 10 u^4)
\nonumber\\
&&{}  + 
     s^3 (t + u)^2 (17 t^4 + 153 t^3 u + 213 t^2 u^2 + 153 t u^3 + 17 u^4)
\nonumber\\
&&{}  + 
     2 s^2 (t + u) (3 t^6 + 42 t^5 u + 122 t^4 u^2 + 168 t^3 u^3 + 122 t^2 u^4
+ 42 t u^5 + 3 u^6)
\nonumber\\
&&{}  + s (t + u)^2 (t^6 + 21 t^5 u + 70 t^4 u^2 + 94 t^3 u^3 + 
       70 t^2 u^4 + 21 t u^5 + u^6)
\nonumber\\
&&{}  + 2 t u (t + u) (t^2 + t u + u^2) (t^4 + 8 t^3 u + 12 t^2 u^2 + 8 t u^3 
+ u^4)]
\nonumber\\
&&{}  + 
   (t + u) [s^8 (t + u) + 
     4 s^7 (t^2 + 3 t u + u^2) + 3 s^6 (t + u) (2 t^2 + 9 t u + 2 u^2)
\nonumber\\
&&{}  + 
     s^5 (4 t^4 + 39 t^3 u + 66 t^2 u^2 + 39 t u^3 + 4 u^4)
\nonumber\\
&&{}  + 
     s^4 (t + u) (t^4 + 24 t^3 u + 36 t^2 u^2 + 24 t u^3 + u^4)
\nonumber\\
&&{}  + 2 s^3 t u (t + u)^2 
      (6 t^2 + 7 t u + 6 u^2) + 
     s^2 t u (t + u) (5 t^4 + 15 t^3 u + 21 t^2 u^2 + 15 t u^3 + 5 u^4)
\nonumber\\
&&{}  + 
     s t u (t + u)^2 (t^4 + 4 t^3 u + 5 t^2 u^2 + 4 t u^3 + u^4) + t^2 u^2 
(t + u) (t^2 + t u + u^2)^2] \} ,
\\
&&\frac{d\sigma}{dt}\left(g g \to Q\overline{Q}\left[
{}^3\!P_J^{(8)}\right]h\right)
=\frac{15}{8} \sum_{J^\prime=0}^{2} (2J^\prime+1)
\frac{d\sigma}{dt}\left(g g \to Q\overline{Q}\left[
{}^3\!P_{J^\prime}^{(1)}\right] h\right),
\\
&&
\frac{d\sigma}{dt}\left(q \overline{q} \to Q\overline{Q}\left[{}^3\!
S_1^{(8)}\right] h\right)
=\frac{4 \pi \alpha_s^2 \left(y_Q^h\right)^2}{27 M s^3 (2 m_h^2 - t - u)^2}
\nonumber\\
&&{}\times
[2 m_h^4 - 2 m_h^2 (3 s + t + u) + 4 s^2 + 4 s (t + u) + t^2 + u^2],
\\
&&\frac{d\sigma}{dt}\left(q \overline{q} \to Q\overline{Q}\left[{}^1\!
P_1^{(8)}\right] h\right)
=\frac{16 \pi \alpha_s^2 \left(y_Q^h\right)^2}{27 M^3 s^3 (2 m_h^2 - t - u)^2}
\nonumber\\
&&{}\times
[2 m_h^4 - 2 m_h^2 (s + t + u) + t^2 + u^2],
\\
&&\frac{d\sigma}{dt}\left(\gamma g \to Q\overline{Q}\left[
{}^1\!S_0^{(8)}\right] h\right)
=\frac{16 \alpha}{\alpha_s}\,
\frac{d\sigma}{dt}\left(g g \to Q\overline{Q}\left[
{}^1\!S_0^{(1)}\right] h\right),
\\
&&\frac{d\sigma}{dt}\left(\gamma g \to Q\overline{Q}\left[
{}^3\!P_J^{(8)}\right] h\right)
=\frac{16 \alpha}{\alpha_s}
\sum_{J^\prime=0}^{2} (2J^\prime+1)
\frac{d\sigma}{dt}\left(g g \to Q\overline{Q}\left[
{}^3\!P_{J^\prime}^{(1)}\right] h\right),
\\
&&\frac{d\sigma}{dt}\left(\gamma \gamma \to Q\overline{Q}\left[
n\right] h\right)
=\frac{512\alpha^2}{9\alpha_s^2}\,
\frac{d\sigma}{dt}\left(g g \to Q\overline{Q}\left[n\right] h\right),
\qquad n={}^1\!S_0^{(1)},{}^3\!P_J^{(1)},
\\
&&\frac{d\sigma}{dt}\left(g g \to Q\overline{Q}\left[{}^1\!
S_0^{(1)}\right] A\right)
=\frac{2 \pi \alpha_s^2 \left(y_Q^A\right)^2 M}
{9 s^2 (m_{A}^2 - s - t)^2 (m_{A}^2 - s - u)^2 
  (2 m_{A}^2 - t - u)^2}
\nonumber\\
&&{}\times\{ m_{A}^8 - 2 m_{A}^6 (s + t + u)
+ m_{A}^4 [2 s^2 + 
     2 s (t + u) + t^2 + 4 t u + u^2]  
- 2 m_{A}^2 t u (s + t + u) 
\nonumber\\
&&{}+ t^2 u^2 \} ,
\\
&&\frac{d\sigma}{dt}\left(g g \to Q\overline{Q}\left[{}^3\!
P_J^{(1)}\right] A\right)
=\frac{-4 \pi \alpha_s^2 \left(y_Q^A\right)^2}
{135 M^3 s^2 (m_{A}^2 - s - t)^4 (m_{A}^2 - s - u)^4 (2 m_{A}^2 - t - u)^4}
\nonumber\\
&&{}\times G_J, \qquad J=0, 1, 2,
\\
&&G_0
= -10 M^2 \{ 16 m_{A}^{16} (t - u)^2 - 
   16 m_{A}^{14} (t - u)^2 [3 s + 4 (t + u)]
\nonumber\\
&&{} + 
   4 m_{A}^{12} [s^4 - 4 s^3 (t + u) + 
     s^2 (17 t^2 - 18 t u + 17 u^2) + 38 s (t + u) (t - u)^2
\nonumber\\
&&{} + 
     2 (t - u)^2 (13 t^2 + 30 t u + 13 u^2)]
\nonumber\\
&&{} - 
   4 m_{A}^{10} [4 s^5 - 12 s^4 (t + u) + 2 s^3 (3 t^2 - 2 t u + 3 u^2) + 
     s^2 (t + u) (47 t^2 - 46 t u + 47 u^2)
\nonumber\\
&&{} + 
     s (t - u)^2 (47 t^2 + 110 t u + 47 u^2) + 
     2 (t + u) (t - u)^2 (11 t^2 + 34 t u + 11 u^2)]
\nonumber\\
&&{} + 
   m_{A}^8 [24 s^6 - 48 s^5 (t + u) - 72 s^4 (t^2 + t u + u^2) + 
     20 s^3 (t + u) (7 t^2 + 2 t u + 7 u^2)
\nonumber\\
&&{} + 
     3 s^2 (71 t^4 + 88 t^3 u + 2 t^2 u^2 + 88 t u^3 + 71 u^4)
%\nonumber\\
%&&{}
+ 6 s (t + u) (t - u)^2 (19 t^2 + 62 t u + 19 u^2)
\nonumber\\
&&{} + 
     (t - u)^2 (41 t^4 + 276 t^3 u + 486 t^2 u^2 + 276 t u^3 + 41 u^4)]
\nonumber\\
&&{} - 
   2 m_{A}^6 [8 s^7 - 8 s^6 (t + u) - 2 s^5 (23 t^2 + 26 t u + 23 u^2)
%\nonumber\\
%&&{}
+ 2 s^4 (t + u) (5 t^2 + 12 t u + 5 u^2)
\nonumber\\
&&{} + 
     s^3 (93 t^4 + 192 t^3 u + 230 t^2 u^2 + 192 t u^3 + 93 u^4)
\nonumber\\
&&{} + s^2 (t + u) (65 t^4 + 96 t^3 u - 2 t^2 u^2 + 96 t u^3 + 65 u^4)
\nonumber\\
&&{} + 
     s (t - u)^2 (17 t^4 + 128 t^3 u + 230 t^2 u^2 + 128 t u^3 + 17 u^4)
\nonumber\\
&&{} + 
     (t + u) (t - u)^2 (5 t^4 + 52 t^3 u + 110 t^2 u^2 + 52 t u^3 + 5 u^4)]
\nonumber\\
&&{} + m_{A}^4 [4 s^8 - 12 s^6 (3 t^2 + 4 t u + 3 u^2) - 
     4 s^5 (t + u) (7 t^2 + 2 t u + 7 u^2)
\nonumber\\
&&{} + 
     2 s^4 (37 t^4 + 96 t^3 u + 136 t^2 u^2 + 96 t u^3 + 37 u^4)
\nonumber\\
&&{} + 2 s^3 (t + u) (57 t^4 + 98 t^3 u + 130 t^2 u^2 + 98 t u^3 + 57 u^4)
\nonumber\\
&&{} + s^2 (47 t^6 + 164 t^5 u + 197 t^4 u^2 + 
       144 t^3 u^3 + 197 t^2 u^4 + 164 t u^5 + 47 u^6)
\nonumber\\
&&{} + 
     4 s (t + u) (t - u)^2 (t^4 + 15 t^3 u + 34 t^2 u^2 + 15 t u^3 + u^4)
\nonumber\\
&&{} + 
     (t^2 - u^2)^2 (t^4 + 22 t^3 u + 66 t^2 u^2 + 22 t u^3 + u^4)]
\nonumber\\
&&{} + 2 m_{A}^2 (t + u) [2 s^7 (t + u) + 
     4 s^6 (t^2 + t u + u^2) - 
     s^5 (t + u) (5 t^2 + 8 t u + 5 u^2)
\nonumber\\
&&{} - 
     s^4 (19 t^4 + 42 t^3 u + 58 t^2 u^2 + 42 t u^3 + 19 u^4)
\nonumber\\
&&{} - 
     s^3 (t + u) (17 t^4 + 23 t^3 u + 36 t^2 u^2 + 23 t u^3 + 17 u^4)
\nonumber\\
&&{} - 
     s^2 (5 t^6 + 15 t^5 u + 21 t^4 u^2 + 14 t^3 u^3 + 21 t^2 u^4 + 
       15 t u^5 + 5 u^6)
\nonumber\\
&&{} - 
     3 s t u (t + u) (t - u)^2 (t^2 + 4 t u + u^2) - t u (t^2 - u^2)^2
(t^2 + 6 t u + u^2)]
\nonumber\\
&&{} + (t + u)^2 (s + t + u)^2 
    [s^4 (t + u)^2 + 
     2 s^3 (t + u) (t^2 + u^2) + s^2 (t^2 + u^2)^2
\nonumber\\
&&{} + t^2 u^2 (t - u)^2] \} ,
\\
&&G_1
= 5 \{ 64 m_{A}^{22} - 48 m_{A}^{20} [9 s + 8 (t + u)]
\nonumber\\
&&{} + 
   4 m_{A}^{18} [325 s^2 + 584 s (t + u) + 4 (63 t^2 + 134 t u + 63 u^2)]
\nonumber\\
&&{} - 
   4 m_{A}^{16} [566 s^3 + 1563 s^2 (t + u) + 
     s (1381 t^2 + 2890 t u + 1381 u^2)
\nonumber\\
&&{} + 
     20 (t + u) (19 t^2 + 46 t u + 19 u^2)]
\nonumber\\
&&{} + 
   4 m_{A}^{14} [619 s^4 + 2376 s^3 (t + u) + s^2 (3267 t^2 + 6758 t u 
+ 3267 u^2)
\nonumber\\
&&{} + 
     9 s (t + u) (209 t^2 + 478 t u + 209 u^2)
\nonumber\\
&&{} + 363 t^4 + 1796 t^3 u + 2882 t^2 u^2 + 1796 t u^3 + 
     363 u^4]
\nonumber\\
&&{} - 
   2 m_{A}^{12} [864 s^5 + 4422 s^4 (t + u) + 
     s^3 (8557 t^2 + 17582 t u + 8557 u^2)
\nonumber\\
&&{} + 
     4 s^2 (t + u) (1951 t^2 + 4298 t u + 1951 u^2)
\nonumber\\
&&{} + 
     s (3269 t^4 + 15024 t^3 u + 23558 t^2 u^2 + 15024 t u^3 + 3269 u^4)
\nonumber\\
&&{} + 
     8 (t + u) (57 t^4 + 332 t^3 u + 566 t^2 u^2 + 332 t u^3 + 57 u^4)]
\nonumber\\
&&{} + m_{A}^{10} [748 s^6 + 5072 s^5 (t + u) + 
     2 s^4 (6631 t^2 + 13590 t u + 6631 u^2)
\nonumber\\
&&{} + 
     2 s^3 (t + u) (8637 t^2 + 18662 t u + 8637 u^2)
\nonumber\\
&&{} + 
     s^2 (11747 t^4 + 51776 t^3 u + 80050 t^2 u^2 + 51776 t u^3 + 
       11747 u^4)
\nonumber\\
&&{} + 4 s (t + u) (946 t^4 + 4773 t^3 u + 7714 t^2 u^2 + 4773 t u^3 + 946 u^4)
\nonumber\\
&&{} + 
     (t + u)^2 (377 t^4 + 2748 t^3 u + 5174 t^2 u^2 + 2748 t u^3 + 377 u^4)]
\nonumber\\
&&{} - 
   m_{A}^8 [184 s^7 + 1716 s^6 (t + u) + 16 s^5 (378 t^2 + 775 t u + 378 u^2)
\nonumber\\
&&{} + 
     4 s^4 (t + u) (2697 t^2 + 5798 t u + 2697 u^2)
\nonumber\\
&&{} + s^3 (10665 t^4 + 46174 t^3 u + 70962 t^2 u^2 + 
       46174 t u^3 + 10665 u^4)
\nonumber\\
&&{} + s^2 (t + u) (5765 t^4 + 27152 t^3 u + 42606 t^2 u^2 + 27152 t u^3 
+ 5765 u^4)
\nonumber\\
&&{} + 
     s (t + u)^2 (1471 t^4 + 8390 t^3 u + 14046 t^2 u^2 + 8390 t u^3 + 
       1471 u^4)
\nonumber\\
&&{} + 
     (t + u)^3 (99 t^4 + 980 t^3 u + 2162 t^2 u^2 + 980 t u^3 + 99 u^4)]
\nonumber\\
&&{} + 
   m_{A}^6 [20 s^8 + 304 s^7 (t + u) + 80 s^6 (19 t^2 + 39 t u + 19 u^2)
\nonumber\\
&&{} + 
     4 s^5 (t + u) (927 t^2 + 2002 t u + 927 u^2)
\nonumber\\
&&{} + 2 s^4 (2549 t^4 + 11026 t^3 u + 16952 t^2 u^2 + 11026 t u^3 + 2549 u^4)
\nonumber\\
&&{} + 8 s^3 (t + u) (512 t^4 + 2357 t^3 u + 3666 t^2 u^2 + 2357 t u^3 
+ 512 u^4)
\nonumber\\
&&{} + 
     s^2 (t + u)^2 (1829 t^4 + 9506 t^3 u + 15006 t^2 u^2 + 9506 t u^3 + 
       1829 u^4)
\nonumber\\
&&{} + 
     2 s (t + u) (187 t^6 + 1640 t^5 u + 4899 t^4 u^2 + 6900 t^3 u^3 + 
  4899 t^2 u^4 + 1640 t u^5 + 187 u^6)
\nonumber\\
&&{} + 
     5 (t + u)^4 (3 t^4 + 46 t^3 u + 130 t^2 u^2 + 46 t u^3 + 3 u^4)]
\nonumber\\
&&{} - 
   m_{A}^4 [20 s^8 (t + u) + 2 s^7 (89 t^2 + 182 t u + 89 u^2)
%\nonumber\\
%&&{}
+ 4 s^6 (t + u) (157 t^2 + 342 t u + 157 u^2)
\nonumber\\
&&{} + 
     4 s^5 (301 t^4 + 1317 t^3 u + 2036 t^2 u^2 + 1317 t u^3 + 301 u^4)
\nonumber\\
&&{} + 
     2 s^4 (t + u) (689 t^4 + 3194 t^3 u + 5012 t^2 u^2 + 3194 t u^3 + 689 u^4)
\nonumber\\
&&{} + s^3 (941 t^6 + 6676 t^5 u + 18033 t^4 u^2 + 24588 t^3 u^3 + 
       18033 t^2 u^4 + 6676 t u^5 + 941 u^6)
\nonumber\\
&&{} + 
     s^2 (t + u) (353 t^6 + 2874 t^5 u + 8015 t^4 u^2 + 11020 t^3 u^3 + 
  8015 t^2 u^4 + 2874 t u^5 + 353 u^6)
\nonumber\\
&&{} + 
     s (57 t^8 + 738 t^7 u + 3246 t^6 u^2 + 7294 t^5 u^3 + 9442 t^4 u^4
\nonumber\\
&&{} + 
       7294 t^3 u^5 + 3246 t^2 u^6 + 738 t u^7 + 57 u^8)
\nonumber\\
&&{} + 
     (t + u)^5 (t^4 + 32 t^3 u + 134 t^2 u^2 + 32 t u^3 + u^4)]
\nonumber\\
&&{} + 
   m_{A}^2 [6 s^8 (t + u)^2 + 
     2 s^7 (t + u) (19 t^2 + 42 t u + 19 u^2)
\nonumber\\
&&{} + 
     s^6 (111 t^4 + 500 t^3 u + 782 t^2 u^2 + 500 t u^3 + 111 u^4)
\nonumber\\
&&{} + 
     4 s^5 (t + u) (47 t^4 + 225 t^3 u + 364 t^2 u^2 + 225 t u^3 + 47 u^4)
\nonumber\\
&&{} + 2 s^4 (96 t^6 + 691 t^5 u + 1909 t^4 u^2 + 2624 t^3 u^3 + 
       1909 t^2 u^4 + 691 t u^5 + 96 u^6)
\nonumber\\
&&{} + 
     2 s^3 (t + u) (57 t^6 + 464 t^5 u + 1294 t^4 u^2 + 1778 t^3 u^3 + 
  1294 t^2 u^4 + 464 t u^5 + 57 u^6)
\nonumber\\
&&{} + 
     s^2 (35 t^8 + 440 t^7 u + 1811 t^6 u^2 + 3888 t^5 u^3 + 4952 t^4 u^4
\nonumber\\
&&{} + 
       3888 t^3 u^5 + 1811 t^2 u^6 + 440 t u^7 + 35 u^8)
\nonumber\\
&&{} + 
     4 s (t + u) (t^8 + 20 t^7 u + 93 t^6 u^2 + 215 t^5 u^3 + 278 t^4 u^4
\nonumber\\
&&{}  + 
  215 t^3 u^5 + 93 t^2 u^6 + 20 t u^7 + u^8) + t u (t + u)^6 (2 t^2 + 17 t u 
+ 2 u^2)]
\nonumber\\
&&{} - 
   (t + u) [s^7 (t + u) (t^2 + 4 t u + u^2) + 
     s^6 (5 t^4 + 28 t^3 u + 50 t^2 u^2 + 28 t u^3 + 5 u^4)
\nonumber\\
&&{} + 
     2 s^5 (t + u) (5 t^4 + 28 t^3 u + 52 t^2 u^2 + 28 t u^3 + 5 u^4)
\nonumber\\
&&{} + 
     2 s^4 (t + u)^2 (5 t^4 + 32 t^3 u + 56 t^2 u^2 + 32 t u^3 + 5 u^4)
\nonumber\\
&&{} + s^3 (t + u) (5 t^6 + 56 t^5 u + 160 t^4 u^2 + 224 t^3 u^3 
+ 160 t^2 u^4 + 
  56 t u^5 + 5 u^6)
\nonumber\\
&&{} + 
     s^2 (t^8 + 24 t^7 u + 103 t^6 u^2 + 224 t^5 u^3 + 284 t^4 u^4 + 
       224 t^3 u^5 + 103 t^2 u^6 + 24 t u^7 + u^8)
\nonumber\\
&&{} + 
     s t u (t + u) (4 t^6 + 19 t^5 u + 46 t^4 u^2 + 58 t^3 u^3 + 46 t^2 u^4 + 
  19 t u^5 + 4 u^6)
\nonumber\\
&&{} + t^2 u^2 (t + u)^6] 
  \} ,
\\
&&G_2
= 384 m_{A}^{22} - 48 m_{A}^{20} [41 s + 48 (t + u)]
\nonumber\\
&&{} + 
   4 m_{A}^{18} [1071 s^2 + 2604 s (t + u) + 4 (379 t^2 + 802 t u + 379 u^2)]
\nonumber\\
&&{} - 
   4 m_{A}^{16} [1350 s^3 + 4905 s^2 (t + u) + 
     s (6001 t^2 + 12634 t u + 6001 u^2)
\nonumber\\
&&{} + 
     20 (t + u) (115 t^2 + 274 t u + 115 u^2)]
\nonumber\\
&&{} + 
   4 m_{A}^{14} [1201 s^4 + 5216 s^3 (t + u) + 
     s^2 (9677 t^2 + 20274 t u + 9677 u^2)
\nonumber\\
&&{} + 
     6 s (t + u) (1319 t^2 + 3082 t u + 1319 u^2)
\nonumber\\
&&{} + 
     4 (555 t^4 + 2696 t^3 u + 4298 t^2 u^2 + 2696 t u^3 + 555 u^4)]
\nonumber\\
&&{} - 
   2 m_{A}^{12} [1768 s^5 + 7602 s^4 (t + u) + 
     s^3 (17081 t^2 + 35646 t u + 17081 u^2)
\nonumber\\
&&{} + 
     8 s^2 (t + u) (2706 t^2 + 6175 t u + 2706 u^2)
\nonumber\\
&&{} + 
     3 s (4399 t^4 + 20772 t^3 u + 32874 t^2 u^2 + 20772 t u^3 + 4399 u^4)
\nonumber\\
&&{} + 
     16 (t + u) (177 t^4 + 1000 t^3 u + 1678 t^2 u^2 + 1000 t u^3 + 177 u^4)]
\nonumber\\
&&{} + m_{A}^{10} [2116 s^6 + 
     8896 s^5 (t + u) + 2 s^4 (9939 t^2 + 20686 t u + 9939 u^2)
\nonumber\\
&&{} + 
     2 s^3 (t + u) (15713 t^2 + 34582 t u + 15713 u^2)
\nonumber\\
&&{} + 
     s^2 (30493 t^4 + 138040 t^3 u + 216542 t^2 u^2 + 138040 t u^3 + 
       30493 u^4)
\nonumber\\
&&{} + 8 s (t + u) (1812 t^4 + 9605 t^3 u + 15890 t^2 u^2 + 9605 t u^3 
+ 1812 u^4)
\nonumber\\
&&{} + 2391 t^6 + 21478 t^5 u + 
     66169 t^4 u^2 + 94100 t^3 u^3 + 66169 t^2 u^4 + 21478 t u^5 + 2391 u^6]
\nonumber\\
&&{} - 
   m_{A}^8 [856 s^7 + 
     4028 s^6 (t + u) + 16 s^5 (556 t^2 + 1153 t u + 556 u^2)
\nonumber\\
&&{} + 
     4 s^4 (t + u) (3631 t^2 + 7486 t u + 3631 u^2)
\nonumber\\
&&{} + 
     s^3 (18407 t^4 + 76658 t^3 u + 118270 t^2 u^2 + 76658 t u^3 + 
       18407 u^4)
\nonumber\\
&&{} + s^2 (t + u) (14251 t^4 + 67500 t^3 u + 110090 t^2 u^2 + 67500 t u^3 + 
  14251 u^4)
\nonumber\\
&&{} + 
     s (5289 t^6 + 43036 t^5 u + 127559 t^4 u^2 + 179784 t^3 u^3
\nonumber\\
&&{} + 
       127559 t^2 u^4 + 43036 t u^5 + 5289 u^6)
\nonumber\\
&&{} + (t + u) (645 t^6 + 7346 t^5 u + 25323 t^4 u^2 + 37052 t^3 u^3
\nonumber\\
&&{} + 
  25323 t^2 u^4 + 7346 t u^5 + 645 u^6)]
\nonumber\\
&&{} + 
   m_{A}^6 [188 s^8 + 1104 s^7 (t + u) + 8 s^6 (357 t^2 + 736 t u + 357 u^2)
\nonumber\\
&&{} + 
     8 s^5 (t + u) (613 t^2 + 1150 t u + 613 u^2)
\nonumber\\
&&{} + 
     s^4 (7042 t^4 + 24892 t^3 u + 36808 t^2 u^2 + 24892 t u^3 + 7042 u^4)
\nonumber\\
&&{} + 
     8 s^3 (t + u) (932 t^4 + 3449 t^3 u + 5458 t^2 u^2 + 3449 t u^3 + 932 u^4)
\nonumber\\
&&{} + 
     s^2 (4531 t^6 + 30636 t^5 u + 84881 t^4 u^2 + 117904 t^3 u^3
\nonumber\\
&&{} + 
       84881 t^2 u^4 + 30636 t u^5 + 4531 u^6)
\nonumber\\
&&{} + 
     6 s (t + u) (209 t^6 + 1962 t^5 u + 6259 t^4 u^2 + 9092 t^3 u^3 + 
  6259 t^2 u^4 + 1962 t u^5 + 209 u^6)
\nonumber\\
&&{} + (t + u)^2 (101 t^6 + 1664 t^5 u + 6795 t^4 u^2 + 
       10240 t^3 u^3 + 6795 t^2 u^4 + 1664 t u^5 + 101 u^6)]
\nonumber\\
&&{} - 
   m_{A}^4 [16 s^9 + 124 s^8 (t + u) + 2 s^7 (229 t^2 + 470 t u + 229 u^2)
\nonumber\\
&&{} + 
     4 s^6 (t + u) (261 t^2 + 446 t u + 261 u^2)
\nonumber\\
&&{} + 
     4 s^5 (464 t^4 + 1325 t^3 u + 1794 t^2 u^2 + 1325 t u^3 + 464 u^4)
\nonumber\\
&&{} + 
     2 s^4 (t + u) (1259 t^4 + 2936 t^3 u + 4152 t^2 u^2 + 2936 t u^3 
+ 1259 u^4)
\nonumber\\
&&{} + 
     s^3 (2151 t^6 + 10188 t^5 u + 23655 t^4 u^2 + 31340 t^3 u^3 + 
       23655 t^2 u^4 + 10188 t u^5 + 2151 u^6)
\nonumber\\
&&{} + 
     s^2 (t + u) (967 t^6 + 6318 t^5 u + 17545 t^4 u^2 + 25076 t^3 u^3
\nonumber\\
&&{}  + 
  17545 t^2 u^4 + 6318 t u^5 + 967 u^6)
\nonumber\\
&&{} + 
     s (179 t^8 + 2446 t^7 u + 11502 t^6 u^2 + 27386 t^5 u^3 + 
       36254 t^4 u^4
\nonumber\\
&&{} + 27386 t^3 u^5 + 11502 t^2 u^6 + 2446 t u^7 + 
       179 u^8)
\nonumber\\
&&{} + (t + u)^3 (7 t^6 + 226 t^5 u + 1225 t^4 u^2 + 
       1884 t^3 u^3 + 1225 t^2 u^4 + 226 t u^5 + 7 u^6)]
\nonumber\\
&&{} + m_{A}^2 [14 s^8 (t + u)^2 + 
     2 s^7 (t + u) (39 t^2 + 58 t u + 39 u^2)
\nonumber\\
&&{} + 
     s^6 (249 t^4 + 580 t^3 u + 674 t^2 u^2 + 580 t u^3 + 249 u^4)
\nonumber\\
&&{} + 
     4 s^5 (t + u) (127 t^4 + 174 t^3 u + 166 t^2 u^2 + 174 t u^3 + 127 u^4)
\nonumber\\
&&{} + 2 s^4 (306 t^6 + 935 t^5 u + 1375 t^4 u^2 + 1480 t^3 u^3 + 
       1375 t^2 u^4 + 935 t u^5 + 306 u^6)
\nonumber\\
&&{} + 
     2 s^3 (t + u) (201 t^6 + 738 t^5 u + 1426 t^4 u^2 + 1854 t^3 u^3 + 
  1426 t^2 u^4 + 738 t u^5 + 201 u^6)
\nonumber\\
&&{} + 
     s^2 (125 t^8 + 1044 t^7 u + 3733 t^6 u^2 + 7996 t^5 u^3 + 
       10328 t^4 u^4
\nonumber\\
&&{} + 7996 t^3 u^5 + 3733 t^2 u^6 + 1044 t u^7 + 125 u^8)
\nonumber\\
&&{} + 
     4 s (t + u) (3 t^8 + 62 t^7 u + 308 t^6 u^2 + 784 t^5 u^3 + 1058 t^4 u^4
\nonumber\\
&&{} + 
  784 t^3 u^5 + 308 t^2 u^6 + 62 t u^7 + 3 u^8)
\nonumber\\
&&{} + 
     t u (t + u)^4 (14 t^4 + 135 t^3 u + 206 t^2 u^2 + 135 t u^3 + 14 u^4)]
\nonumber\\
&&{} - 
   (t + u) [ 
     s^7 (t + u) (7 t^2 - 4 t u + 7 u^2) + 
     s^6 (35 t^4 + 28 t^3 u - 2 t^2 u^2 + 28 t u^3 + 35 u^4)
\nonumber\\
&&{} + 
     2 s^5 (t + u) (35 t^4 + 12 t^3 u - 8 t^2 u^2 + 12 t u^3 + 35 u^4)
\nonumber\\
&&{} + 2 s^4 (t + u)^2 
      (35 t^4 + 8 t^3 u + 6 t^2 u^2 + 8 t u^3 + 35 u^4)
\nonumber\\
&&{} + 
     s^3 (t + u) (35 t^6 + 104 t^5 u + 132 t^4 u^2 + 176 t^3 u^3 
+ 132 t^2 u^4 + 
  104 t u^5 + 35 u^6)
\nonumber\\
&&{} + 
     s^2 (7 t^8 + 64 t^7 u + 197 t^6 u^2 + 420 t^5 u^3 + 548 t^4 u^4 + 
       420 t^3 u^5 + 197 t^2 u^6 + 64 t u^7 + 7 u^8)
\nonumber\\
&&{} + 
     3 s t u (t + u) (4 t^6 + 19 t^5 u + 54 t^4 u^2 + 74 t^3 u^3 + 54 t^2 u^4 + 
  19 t u^5 + 4 u^6)
\nonumber\\
&&{} + t^2 u^2 (t + u)^4 (7 t^2 + 10 t u + 7 u^2) ] ,
\\
&&\frac{d\sigma}{dt}\left(g g \to Q\overline{Q}\left[
{}^1\!S_0^{(8)}\right]A\right)
=\frac{15}{8}\,
\frac{d\sigma}{dt}\left(g g \to Q\overline{Q}\left[
{}^1\!S_0^{(1)}\right] A \right),
\\
&&\frac{d\sigma}{dt}\left(g g \to Q\overline{Q}\left[{}^3\!
S_1^{(8)}\right] A\right)
=
\frac{\pi \alpha_s^2 \left(y_Q^A\right)^2}
{4 M s^3 (m_{A}^2 - s - t)^2 (m_{A}^2 - s - u)^2 (2 m_{A}^2 - t - u)^2} 
\nonumber\\
&&{}\times \{ 9 m_{A}^{12} - m_{A}^{10} [20 s + 27 (t + u)]
%\nonumber\\
%&&{}
+ m_{A}^8 [14 s^2 + 46 s (t + u) + 3 (11 t^2 + 23 t u + 11 u^2)]
\nonumber\\
&&{} - 
   m_{A}^6 [3 s^3 + 23 s^2 (t + u) + 42 s (t + u)^2 + 
     3 (t + u) (7 t^2 + 16 t u + 7 u^2)]
\nonumber\\
&&{} + 
   m_{A}^4 [3 s^3 (t + u) + 
     2 s^2 (7 t^2 + 13 t u + 7 u^2) + 2 s (t + u) (9 t^2 + 20 t u + 9 u^2)
\nonumber\\
&&{} + 7 t^4 + 35 t^3 u + 51 t^2 u^2 + 35 t u^3 + 7 u^4]
\nonumber\\
&&{} - m_{A}^2 [s^3 (t^2 + t u + u^2) + s^2 (t + u) (3 t^2 + 7 t u + 3 u^2)
\nonumber\\
&&{} + s (3 t^4 + 18 t^3 u + 26 t^2 u^2 + 18 t u^3 + 3 u^4) + (t + u)
(t^2 + t u + u^2) (t^2 + 7 t u + u^2)]
\nonumber\\
&&{} + 
   t u [s (t + u) + t^2 + t u + u^2]^2
 \} ,
\\
&&\frac{d\sigma}{dt}\left(g g \to Q\overline{Q}\left[{}^1\!
P_1^{(8)}\right] A\right)
=
\frac{\pi \alpha_s^2 \left(y_Q^A\right)^2}
{M^3 s^3 (m_{A}^2 - s - t)^3 (m_{A}^2 - s - u)^3 (2 m_{A}^2 - t - u)^4}
\nonumber\\
&&{} \times\{ 3 m_{A}^{16} [4 s^2 + 4 s (t + u) + 3 (t +
u)^2]
\nonumber\\
&&{}  - 
   m_{A}^{14} [28 s^3 + 76 s^2 (t + u) + 
     s (73 t^2 + 142 t u + 73 u^2) + 36 (t + u)^3]
\nonumber\\
&&{}  + 
   m_{A}^{12} [28 s^4 + 108 s^3 (t + u) + 
     2 s^2 (97 t^2 + 204 t u + 97 u^2)
\nonumber\\
&&{}  + s (t + u) (161 t^2 + 330 t u + 161 u^2) + 12 (t + u)^2 (5 t^2 + 11 t u
+ 5 u^2)]
\nonumber\\
&&{}  - m_{A}^{10} [20 s^5 + 76 s^4 (t + u) + 
     4 s^3 (43 t^2 + 97 t u + 43 u^2)
\nonumber\\
&&{}  + 2 s^2 (t + u) (127 t^2 + 302 t u + 127 u^2)
\nonumber\\
&&{}  + s (181 t^4 + 783 t^3 u + 1192 t^2 u^2 + 
       783 t u^3 + 181 u^4) + 
     18 (t + u)^3 (3 t^2 + 8 t u + 3 u^2)]
\nonumber\\
&&{}  + 
   m_{A}^8 [8 s^6 + 40 s^5 (t + u) + s^4 (79 t^2 + 190 t u + 79 u^2)
\nonumber\\
&&{}  + 
     2 s^3 (t + u) (73 t^2 + 206 t u + 73 u^2) + 
     4 s^2 (48 t^4 + 229 t^3 u + 363 t^2 u^2 + 229 t u^3 + 48 u^4)
\nonumber\\
&&{}  + 
     s (t + u) (117 t^4 + 560 t^3 u + 866 t^2 u^2 + 560 t u^3 + 117 u^4)
\nonumber\\
&&{}  + 
     2 (t + u)^2 (14 t^4 + 79 t^3 u + 129 t^2 u^2 + 79 t u^3 + 14 u^4)]
\nonumber\\
&&{}  - m_{A}^6 [8 s^6 (t + u) + s^5 (23 t^2 + 58 t u + 23 u^2) + 
     2 s^4 (t + u) (19 t^2 + 66 t u + 19 u^2)
\nonumber\\
&&{}  + 
     2 s^3 (38 t^4 + 201 t^3 u + 340 t^2 u^2 + 201 t u^3 + 38 u^4)
\nonumber\\
&&{}  + 
     2 s^2 (t + u) (45 t^4 + 226 t^3 u + 384 t^2 u^2 + 226 t u^3 + 45 u^4)
\nonumber\\
&&{}  + s (45 t^6 + 337 t^5 u + 934 t^4 u^2 + 1272 t^3 u^3 + 
       934 t^2 u^4 + 337 t u^5 + 45 u^6)
\nonumber\\
&&{}  + 
     4 (t + u)^3 (2 t^4 + 16 t^3 u + 27 t^2 u^2 + 16 t u^3 + 2 u^4)]
\nonumber\\
&&{}  + 
   m_{A}^4 [2 s^6 (t + u)^2 + s^5 (t + u) (5 t^2 + 14 t u + 5 u^2)
\nonumber\\
&&{}  + 
     2 s^4 (7 t^4 + 31 t^3 u + 58 t^2 u^2 + 31 t u^3 + 7 u^4)
\nonumber\\
&&{}  + 
     s^3 (t + u) (29 t^4 + 122 t^3 u + 262 t^2 u^2 + 122 t u^3 + 29 u^4)
\nonumber\\
&&{}  + s^2 (27 t^6 + 180 t^5 u + 539 t^4 u^2 + 
       768 t^3 u^3 + 539 t^2 u^4 + 180 t u^5 + 27 u^6)
\nonumber\\
&&{}  + 
     s (t + u) (10 t^6 + 86 t^5 u + 255 t^4 u^2 + 342 t^3 u^3 + 255 t^2 u^4 + 
  86 t u^5 + 10 u^6)
\nonumber\\
&&{}  + (t + u)^2 (t^6 + 18 t^5 u + 63 t^4 u^2 + 88 t^3 u^3 + 
       63 t^2 u^4 + 18 t u^5 + u^6)]
\nonumber\\
&&{}  - 
   m_{A}^2 [s^5 (t^4 + t^3 u + t u^3 + u^4) + 
     4 s^4 (t + u) (t^4 + 2 t^3 u + 5 t^2 u^2 + 2 t u^3 + u^4)
\nonumber\\
&&{}  + 
     2 s^3 (3 t^6 + 16 t^5 u + 48 t^4 u^2 + 76 t^3 u^3 + 48 t^2 u^4 + 
       16 t u^5 + 3 u^6)
\nonumber\\
&&{}  + 2 s^2 (t + u) (2 t^6 + 15 t^5 u + 43 t^4 u^2 + 72 t^3 u^3 + 43 t^2 u^4
+ 15 t u^5 + 2 u^6)
\nonumber\\
&&{}  + 
     s (t^8 + 15 t^7 u + 64 t^6 u^2 + 142 t^5 u^3 + 180 t^4 u^4 + 
       142 t^3 u^5 + 64 t^2 u^6 + 15 t u^7 + u^8)
\nonumber\\
&&{}  + 2 t u (t + u)^3 
      (t^4 + 5 t^3 u + 6 t^2 u^2 + 5 t u^3 + u^4)]
\nonumber\\
&&{}  + t u (t + u) [s^4 (t + u) (t^2 + t u + u^2) + 
     s^3 (3 t^4 + 6 t^3 u + 14 t^2 u^2 + 6 t u^3 + 3 u^4)
\nonumber\\
&&{}  + 
     s^2 (t + u) (3 t^4 + 4 t^3 u + 14 t^2 u^2 + 4 t u^3 + 3 u^4)
\nonumber\\
&&{}  + s (t + u)^2 
      (t^4 + 2 t^3 u + 4 t^2 u^2 + 2 t u^3 + u^4) + t u (t + u) (t^2 + t u 
+ u^2)^2]\},
\\
&&\frac{d\sigma}{dt}\left(g g \to Q\overline{Q}\left[
{}^3\!P_J^{(8)}\right]
A\right)=\frac{15}{8}\,
\sum_{J^\prime=0}^{2}(2J^\prime+1)
\frac{d\sigma}{dt}\left(g g \to Q\overline{Q}\left[
{}^3\!P_{J^\prime}^{(1)}\right] A \right),
\\
&&\frac{d\sigma}{dt}\left(q \overline{q} \to Q\overline{Q}\left[{}^3\!
S_1^{(8)}\right] A\right)
=\frac{4 \pi \alpha_s^2 \left(y_Q^A\right)^2}{27 M s^3 (2 m_{A}^2 - t - u)^2}
\nonumber\\
&&{}\times
[2 m_{A}^4 - 2 m_{A}^2 (s + t + u) + t^2 + u^2],
\\
&&\frac{d\sigma}{dt}\left(q \overline{q} \to Q\overline{Q}\left[{}^1\!
P_1^{(8)}\right] A\right)
=\frac{-16 \pi \alpha_s^2 \left(y_Q^A\right)^2}
{27 M^3 s^3 (2 m_{A}^2 - t - u)^4}
\nonumber\\
&&{}\times
\{ 8 m_{A}^6 s - 2 m_{A}^4 (2 s + t + u)^2 - 2 m_{A}^2 [s (t - u)^2 
- (t + u)^3] - (t + u)^2 (t^2 + u^2)\} ,
\\
&&\frac{d\sigma}{dt}\left(\gamma g \to Q\overline{Q}\left[
{}^1\!S_0^{(8)}\right] A\right)
=\frac{16\alpha}{\alpha_s}\,
\frac{d\sigma}{dt}\left(g g \to Q\overline{Q}\left[
{}^1\!S_0^{(1)}\right] A\right),
\\
&&\frac{d\sigma}{dt}\left(\gamma g \to Q\overline{Q}\left[
{}^3\!P_J^{(8)}\right]A\right)
=\frac{16\alpha}{\alpha_s}\sum_{J^\prime=0}^{2}
(2J^\prime+1)\frac{d\sigma}{dt}\left(g g \to Q\overline{Q}\left[
{}^3\!P_{J^\prime}^{(1)}\right] A\right),
\\
&&\frac{d\sigma}{dt}\left(\gamma \gamma \to Q\overline{Q}\left[
n\right] A\right)
=\frac{512\alpha^2}{9\alpha_s^2}\,
\frac{d\sigma}{dt}\left(g g \to Q\overline{Q}\left[
n\right] A\right),\qquad n={}^1\!S_0^{(1)},{}^3\!P_J^{(1)}.
\end{eqnarray}

\end{appendix}

\newpage
\begin{figure}[ht]
\begin{center}
\epsfig{figure=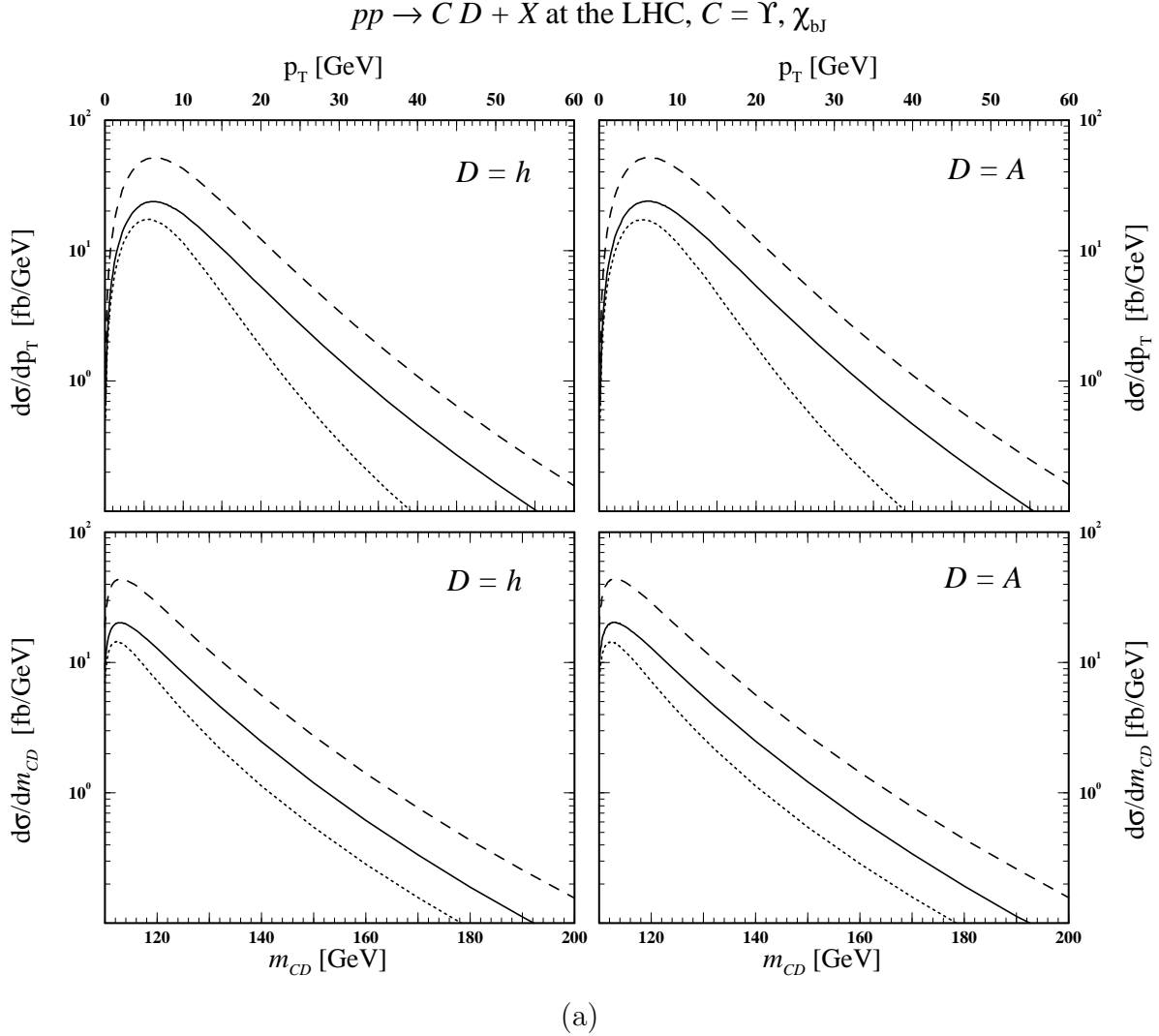,width=\textwidth}
(a)
\caption{(a) $p_T$ distributions $d\sigma/dp_T$ (upper panel) and
$m_{CD}$ distributions $d\sigma/dm_{CD}$ (lower panel) in fb/GeV,
(b) $y_C$ distributions $d\sigma/dy_C$ (upper panel) and $y_D$ distributions
$d\sigma/dy_D$ (lower panel) in fb, and
(c) total cross sections $\sigma$ in fb as functions of $\tan\beta$ (upper
panel) and $m_D$ (lower panel) of $pp\to CD+X$, where
$C=\Upsilon(1S),\chi_{bJ}(1P)$ and $D=h$ (left columns) or $D=A$ (right
columns), at the LHC.
The default values of the MSSM input parameters are $\tan\beta=50$ and 
$m_D=100$~GeV.
It is summed over $C=\chi_{b0}(1P),\chi_{b1}(1P),\chi_{b2}(1P)$.
In each frame, the NRQCD (solid line) prediction for $C=\Upsilon(1S)$ and the
NRQCD (dashed line) and CSM (dotted line) ones for $C=\chi_{bJ}(1P)$ are shown
separately.
\label{fig:l}}
\end{center}
\end{figure}

\newpage
\begin{figure}[ht]
\begin{center}
\epsfig{figure=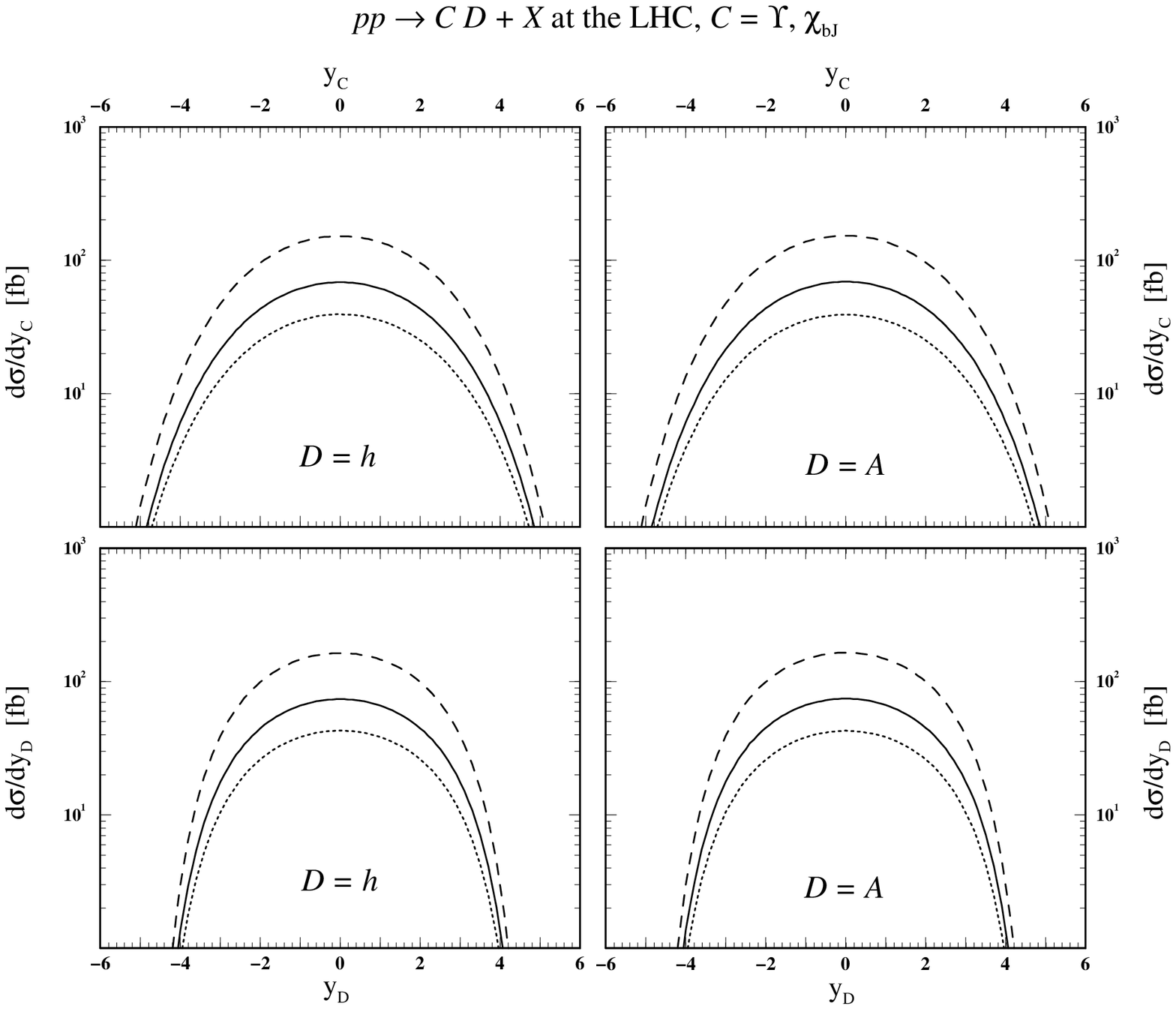,width=\textwidth}
(b)\\
Fig.~\ref{fig:l} (continued).
\end{center}
\end{figure}

\newpage
\begin{figure}[ht]
\begin{center}
\epsfig{figure=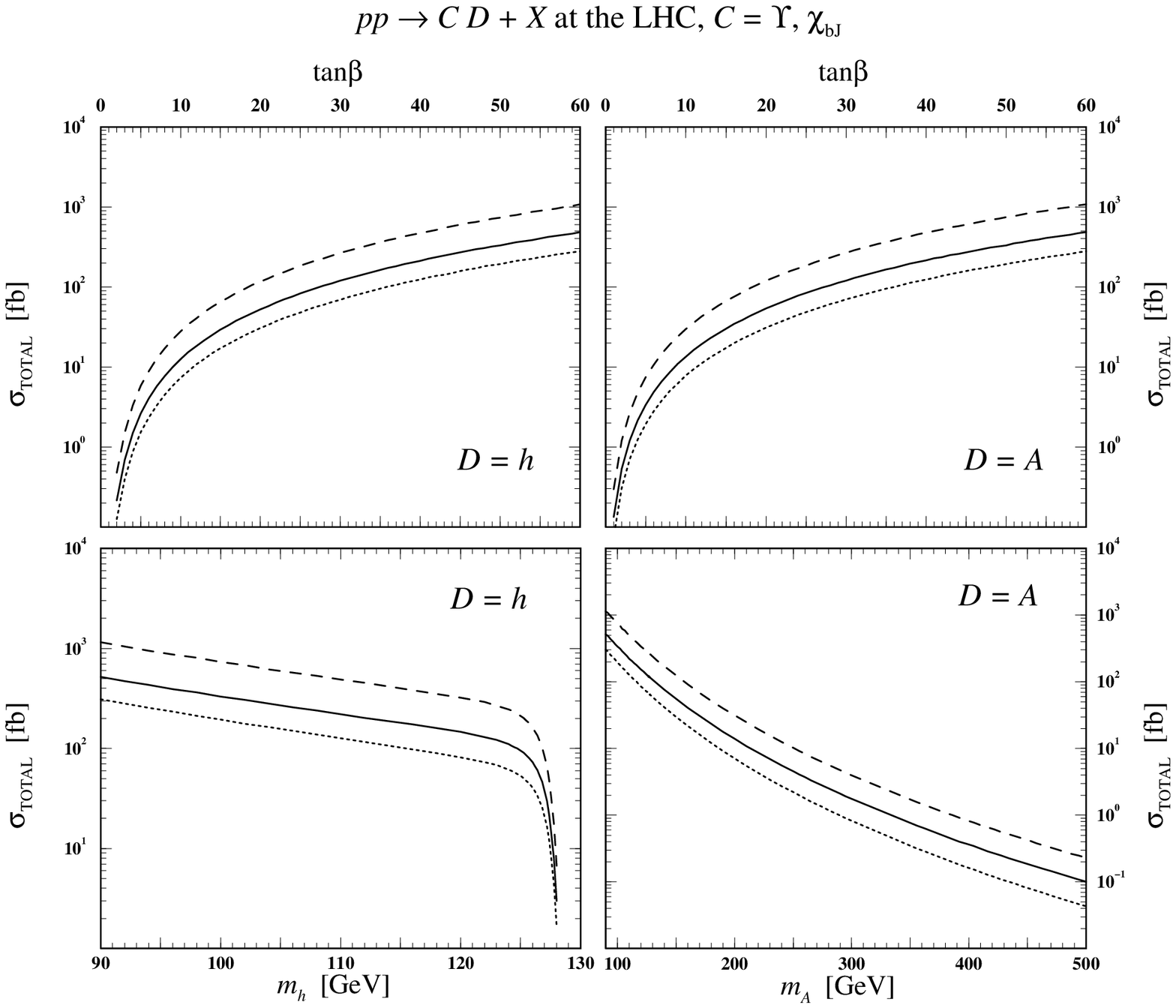,width=\textwidth}
(c)\\
Fig.~\ref{fig:l} (continued).
\end{center}
\end{figure}

\newpage
\begin{figure}[ht]
\begin{center}
\epsfig{figure=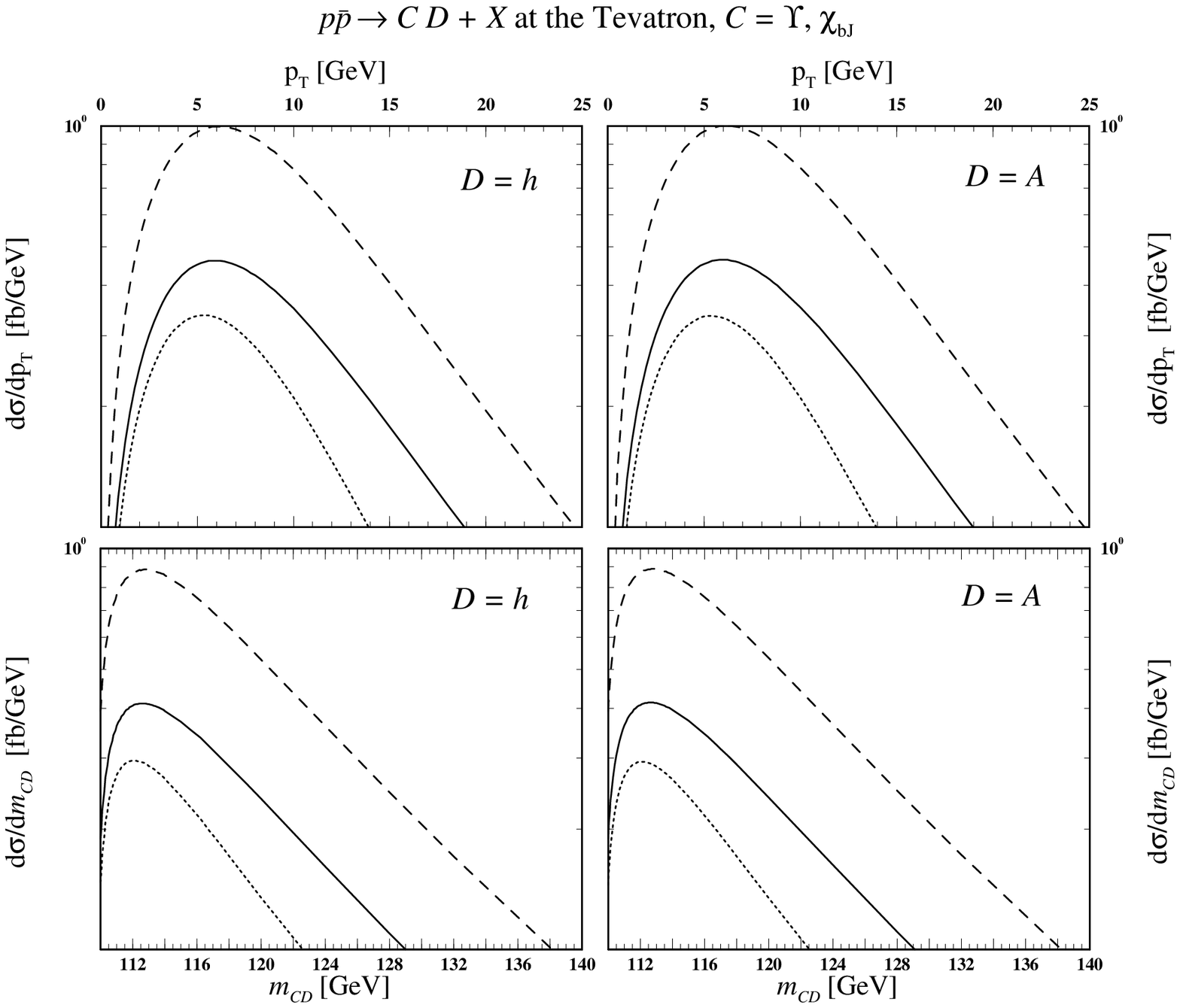,width=\textwidth}
(a)\\
\caption{Same as in Figs.~\ref{fig:l}(a)--(c), but for $p\overline{p}\to CD+X$
in Run~II at the Tevatron.
\label{fig:t}}
\end{center}
\end{figure}

\newpage
\begin{figure}[ht]
\begin{center}
\epsfig{figure=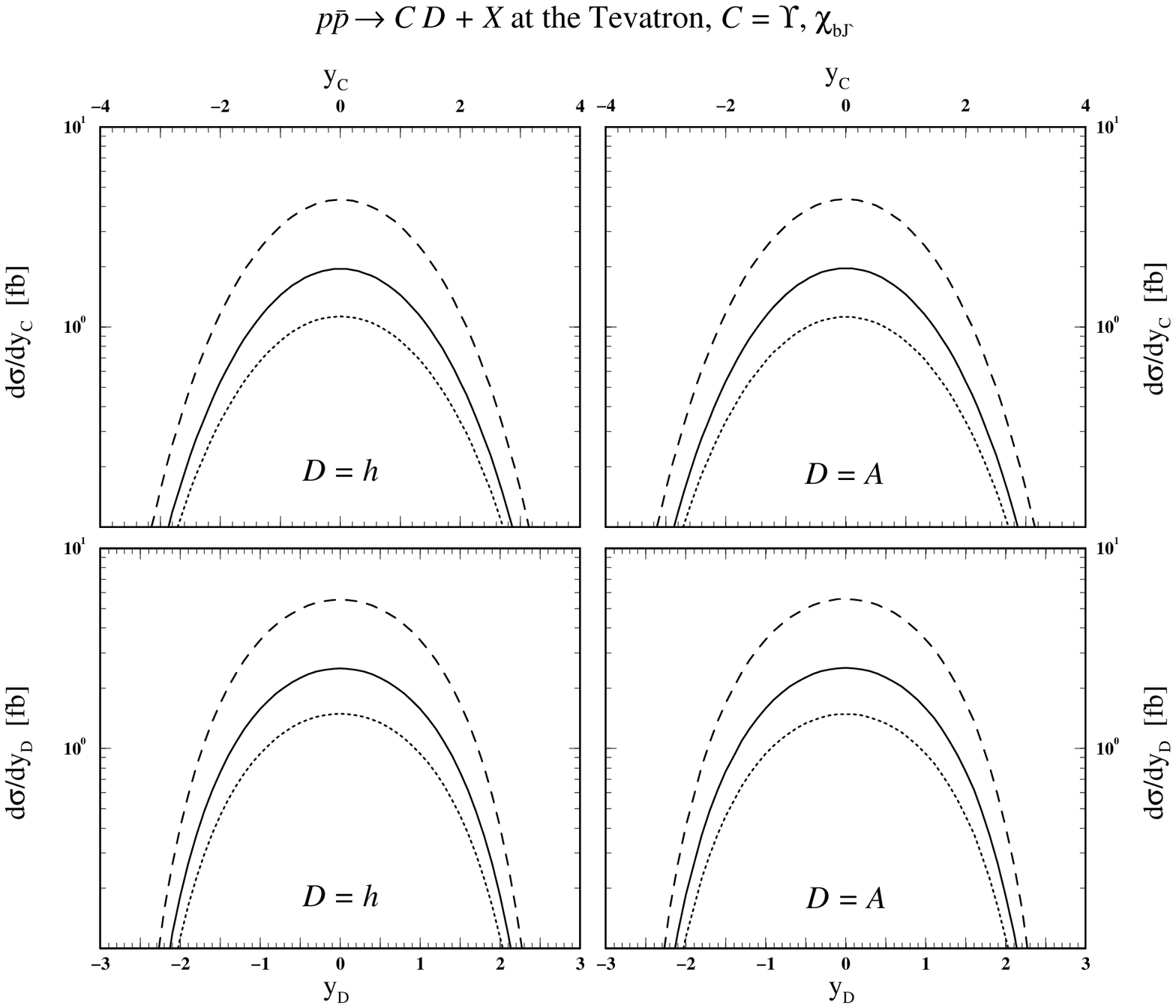,width=\textwidth}
(b)\\
Fig.~\ref{fig:t} (continued).
\end{center}
\end{figure}

\newpage
\begin{figure}[ht]
\begin{center}
\epsfig{figure=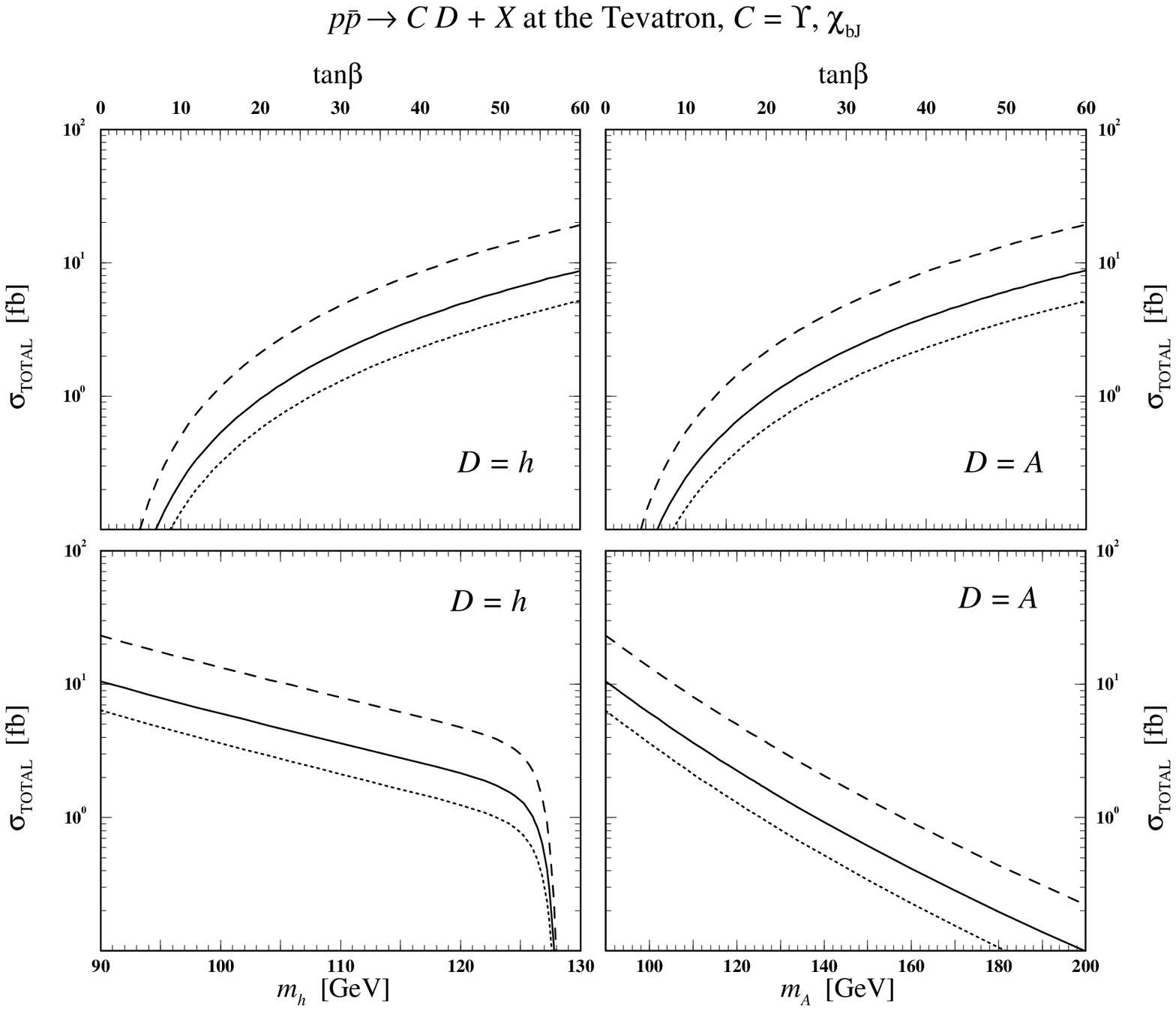,width=\textwidth}
(c)\\
Fig.~\ref{fig:t} (continued).
\end{center}
\end{figure}

\newpage
\begin{figure}[ht]
\begin{center}
\epsfig{figure=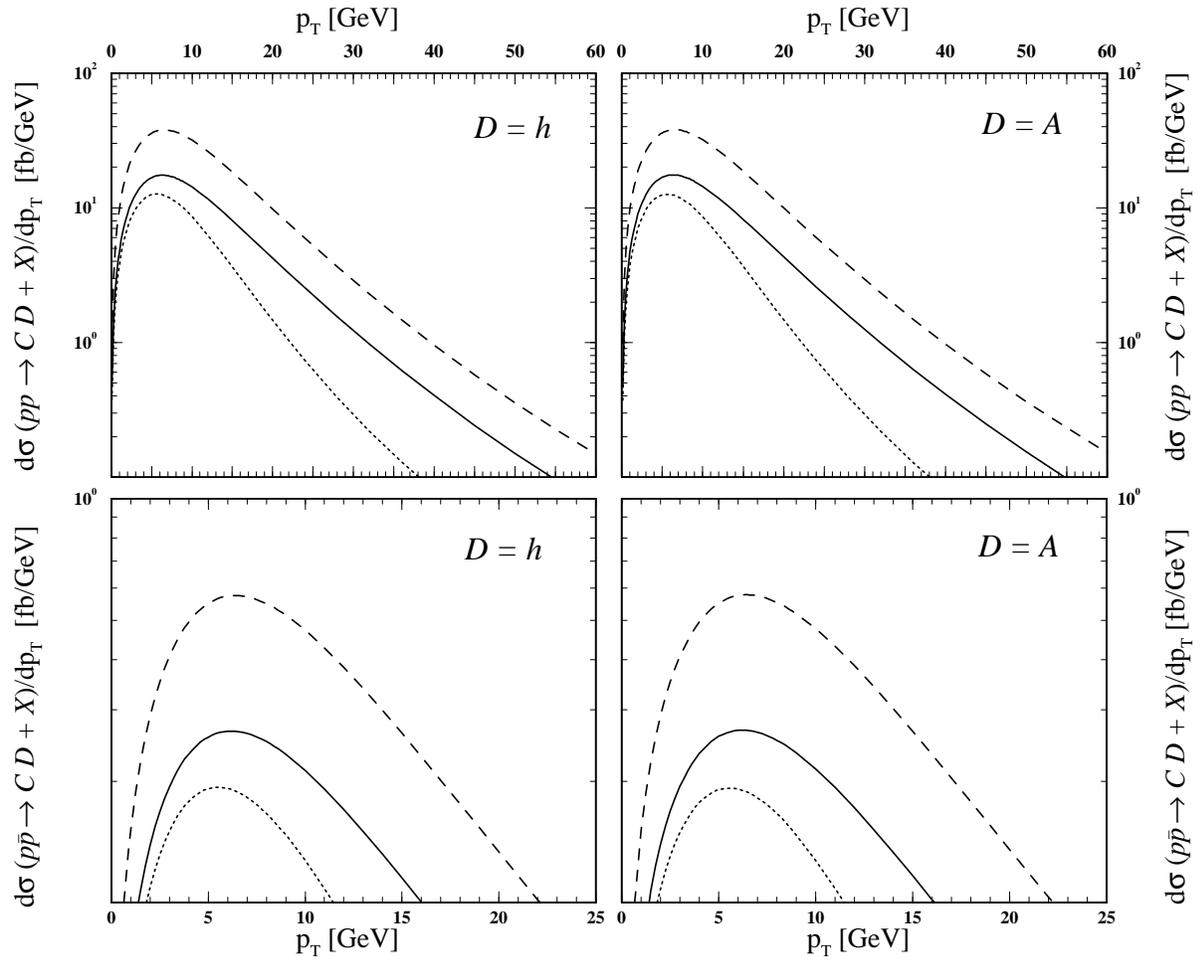,width=\textwidth}
\caption{Same as in the upper panels of Figs.~\ref{fig:l}(a) and
Figs.~\ref{fig:t}(a), but for $\mu_r=\mu_f=m_D$.
\label{fig:s}}
\end{center}
\end{figure}

\end{document}